\documentclass[useAMS,usenatbib]{mn2e}
\usepackage{amssymb,amsmath,graphicx}
 \usepackage{threeparttable}
\usepackage{url}
\usepackage{appendix}

\newcommand{\oiii}[1]{[{\ensuremath{\mathrm{O}}}\,\textsc{iii}]\:#1}
\newcommand{\oii}[1]{[{\ensuremath{\mathrm{O}}}\,\textsc{ii}]\:#1}

\newcommand{\nii}[1]{[{\ensuremath{\mathrm{N}}}\,\textsc{ii}]\:#1}

\newcommand{\sii}[1]{[{\ensuremath{\mathrm{S}}}\,\textsc{ii}]\:#1}

\newcommand{\heii}[1]{{\ensuremath{\mathrm{He}}}\,\textsc{ii}\:#1}

\newcommand{\hii}{{\ensuremath{\mathrm{H}}}\,\textsc{ii}}
\newcommand{\hi}{{\ensuremath{\mathrm{H}}}\,\textsc{i}}

\newcommand{\hb}{\ensuremath{\mathrm{H}\beta}}
\newcommand{\ha}{\ensuremath{\mathrm{H}\alpha}}

\newcommand{\hg}{\ensuremath{\mathrm{H}\gamma}}
\newcommand{\hd}{\ensuremath{\mathrm{H}\delta}}

\usepackage{color}

\usepackage[T1]{fontenc}
\usepackage{ae,aecompl}
\usepackage{hyperref}

\begin{document}

\title[SINFONI observations of the 8 o'clock arc]{The physical nature of the 8 o'clock arc based on near-IR IFU spectroscopy with SINFONI\thanks{Based on observations obtained at the Very Large Telescope (VLT) of the 
European Southern Observatory, Paranal, Chile (ESO Programme ID: 83A:0879A, PI: J. Brinchmann). 
Also based on observations with the NASA/ESA {\it Hubble Space Telescope (HST)}, obtained at the 
Space Telescope Science Institute under the programme ID No. 11167 (PI: S. Allam).}} 
\author[M. Shirazi et al.] 
{M. Shirazi$^1$\thanks{shirazi@strw.leidenuniv.nl}, S. Vegetti$^{2,3}$, N. Nesvadba$^4$, S. Allam$^{5,6}$, J. Brinchmann$^1$, D. Tucker$^5$\\
$^1$ Leiden Observatory, Leiden University, PO Box 9513, 2300 RA Leiden, The Netherlands\\
$^2$ Kavli Institute for Astrophysics and Space Research,Massachusetts Institute of Technology, Cambridge, MA 02139, USA\\
$^3$ Max-Planck Institute for Astrophysics, Karl-Schwarzschild-Strasse 1, D-85740 Garching, Germany\\
$^4$ Institut d'Astrophysique Spatiale, UMR 8617, CNRS, Universit\'{e}  Paris-Sud, B\^{a}timent 121, 91405, Orsay Cedex, France\\
$^5$ Fermi National Accelerator Laboratory, PO Box 500, Batavia, IL 60510, USA\\
$^6$ Space Telescope Science Institute, 3700 San Martin Drive, Baltimore, MD 21218, USA}

\maketitle

\begin{abstract}
We present an analysis of near-infrared integral field unit spectroscopy of the 8 o'clock arc, a gravitationally 
lensed Lyman break galaxy, taken with SINFONI. 
We explore the shape of the spatially-resolved \hb\ profile and demonstrate 
that we can decompose it into three components that partially overlap (spatially) but are 
distinguishable when we include dynamical information. We use existing $B$ and $H$ imaging from the \textit{Hubble Space Telescope} to construct a rigorous lens model using a Bayesian grid based lens 
modelling technique. We apply this lens model to the SINFONI data cube to construct the de-lensed 
\hb\ line-flux velocity and velocity dispersion maps of the galaxy. We find that the 8 o'clock arc has a complex velocity field that is not simply 
explained by a single rotating disk. 
The \hb\ profile of the galaxy shows a blue-shifted wing suggesting gas outflows of $\sim$ 200 $\rm{km\; s^{-1}}$. 
We confirm that the 8 o'clock arc lies on the stellar mass--oxygen abundance--SFR plane found locally, but it has nevertheless significantly different gas surface density (a factor of 2--4 higher) and electron density in the ionized gas (five times higher) from those in similar nearby galaxies, possibly indicating a higher density interstellar medium for this galaxy.

\end{abstract}



\section{Introduction}

The last decade has seen a dramatic increase in our knowledge of the galaxy population at redshift $z \;> 2$. 
In particular, the large samples of high redshift (high-$z$) galaxies identified by the Lyman-break dropout technique 
\citep[][ and references therein]{steidel03} have allowed detailed statistical analysis of the physical properties 
of these galaxies \citep{shapley11}. While early studies made use of long-slit near-infrared (NIR) spectroscopy 
\citep{erb06a, erb06b, erb06c} to study the physical properties of these galaxies, 
more recent studies have focused on NIR integral field units \citep[IFUs;][]{forster06, forster09, genzel08, genzel10}.

The steadily growing effort to obtain resolved NIR spectra of high-$z$ galaxies in a
systematic manner as in the MASSIV, SINS, SINS/zC-SINF and LSD/AMAZE surveys is leading to
samples of spatially-resolved emission line maps of distant ($z\sim 1-3.8$) star-forming galaxies. 
Studying these maps has provided us with spatially-resolved physical properties, metallicity 
gradients and kinematics of high-$z$ star-forming galaxies \citep[e.g.,][]{contini12,epinat12,forster09,cresci09,genzel11,forster11a,forster11b,newman12b,maiolino08,mannucci09,gnerucci11}.

A particularly important question for these studies is whether the observed dynamics are due to, or significantly
influenced by major mergers. While this is generally difficult to establish, \cite{genzel06} have shown that with
sufficiently high resolution integral field unit (IFU) spectroscopy, it is possible to distinguish between rotation and merging. 
However, variations in spatial resolution still cause inconclusive interpretations. 
As an example, using SINFONI observations of 14 Lyman break galaxies (LBGs) \cite{forster06} argued for rotationally
supported dynamics in many LBGs (seven out of nine resolved velocity fields). In contrast, by studying spatially resolved spectra 
of three galaxies at redshift $z\sim 2-3$, using the OSIRIS in combination with adaptive optics (AO), \cite{law07} showed that the ionized 
gas kinematics of those galaxies were inconsistent with simple rotational support.

Analysis of the SINS \ha\ sample studied by \cite{forster09} showed that about one-third of 62 galaxies in their sample 
show rotation-dominated kinematics, another one-third are dispersion-dominated objects, and the remaining galaxies 
are interacting or merging systems. However, more recent AO data have shown that many of these dispersion-dominated 
sources are in fact rotating and follow the same scaling relations as more massive galaxies \citep{newman13}. They also 
show that the ratio of rotation to random motions ($\rm{V}/\sigma$) increases with stellar mass. This result shows the importance of 
spatial resolution for studying high-$z$ galaxies.

While we are essentially limited by the intrinsic faintness of these objects, gravitational lensing can 
significantly magnify these galaxies and allow us to study their properties at a level similar to what is 
achieved at lower redshifts \citep[e.g., MS 1512-cB58; see][]{yee96, pettini00,pettini02,teplitz00, savaglio02,siana08}.
Although NIR IFU observations with AO have been able to spatially resolve high-$z$ galaxies
\citep{forster06, forster09}, obtaining a resolution better than 0.2 arc sec even with AO is very difficult and lensing is the only 
way to obtain sub-kpc scale resolution for high-$z$ galaxies using current instruments. Studies of this nature will truly 
come into their own in the future with 30m-class telescopes.

Given a sufficiently strongly lensed LBG, we might be able to study its dynamical state, 
the influence of any potential non-thermal ionizing source, such as a faint active galactic nucleus (AGN), and 
the physical properties of the interstellar medium (ISM).

Spatially-resolved studies of six strongly lensed star-forming galaxies at $z\sim1.7-3.1$ using the Keck laser 
guide star AO system and the OSIRIS IFU spectrograph enabled \cite{jones10a} to resolve the kinematics of these 
galaxies on sub-kpc scales. Four of these six galaxies display coherent velocity fields consistent with a simple 
rotating disk model. Using the same instrument, \cite{jones10b} also studied spatially-resolved spectroscopy 
of the \textit{Clone arc} in detail. Deriving a steep metallicity gradient for this lensed galaxy at $z=2$, they 
suggested an inside-out assembly history with radial mixing and enrichment from star formation.
A detailed study of the spatially-resolved kinematics for a highly amplified galaxy at $z = 4.92$ by \cite{swinbank09}
suggests that this young galaxy is undergoing its first major epoch of mass assembly. Furthermore, analysing 
NIR spectroscopy for a sample of 28 gravitationally-lensed star-forming galaxies in the redshift 
range $1.5 < z < 5$, observed mostly with the Keck II telescope, \cite{richard11} provided us with the properties 
of a representative sample of low luminosity galaxies at high-$z$.

The small number of bright ${z}\sim 2$ lensed galaxies has recently been increased by a spectroscopic campaign following-up galaxy-galaxy lens candidates within the Sloan Digital Sky Survey \citep[SDSS;][]{stark13}. These high spatial and spectral resolution data, will provide us with constraints on the outflow, metallicity gradients, and stellar populations in high-$z$ galaxies.

Given its interesting configuration and brightness, the 8 o'clock arc \citep{allam07} is of major interest for the detailed
investigation of the physical and kinematical properties of LBGs. Indeed, there has been a vigorous
campaign to obtain a significant collection of data for this object. In particular,  the following observations have been made: 
five-band $Hubble \;Space \;Telescope \;(HST)$ imaging covering $F450W$ to $F160W$, a Keck LRIS spectrum of the rest-frame UV, 
NIR $H$- and $K$-band long-slit spectroscopy with the Near InfraRed Imager and Spectrometer on the 
Gemini North 8m telescope \citep{finkelstein09} and X-shooter observations with the UV-B, VIS-R and NIR spectrograph 
arms \citep[][hereafter DZ10 and DZ11]{DZ10, DZ11}.

Measuring the differences between the redshift of stellar photospheric lines and ISM absorption lines, \cite{finkelstein09} 
suggested gas outflows of the order of 160 $\rm{km\; s^{-1}}$ for this galaxy. DZ10 also showed that the ISM lines are extended over 
a large velocity range up to $\sim$ 800 $\rm{km\; s^{-1}}$ relative to the systematic redshift. They showed that the peak
optical depth of the ISM lines is blue-shifted relative to the stellar photospheric lines, implying gas outflows of 120 $\rm{km\; s^{-1}}$.

Studying the rest frame UV, DZ10 showed that the Ly$\alpha$ line is dominated by a damped absorption 
profile with a weak emission profile redshifted relative to the ISM lines by about $+690$ $\rm{km\; s^{-1}}$ on top of the 
absorption profile. They suggested that this results from backscattered Ly$\alpha$ photons 
emitted in the \hii\ region surrounded by the cold, expanding ISM shell.

DZ11 argued that the 8 o'clock arc is formed of two major parts, the main galaxy component and a smaller 
clump which is rotating around the main core of the galaxy and separated by 1.2 kpc in projected distance. 
They found that the properties of the clump resemble those of the high-$z$ clumps studied by \citet{swinbank09}, 
\citet{jones10a}, and \citet{genzel11}. They also suggested that the fundamental relation between mass, star formation rate (SFR), and metallicity \citep{lara-lopez10,mannucci10}
may hold up to and even beyond $z = 2.5$, as also supported by two other lensed LBGs at $2.5 < z < 3.5$ studied by \cite{richard11}.

In this work, we use NIR IFU spectroscopy of the 8 o'clock arc with SINFONI to spatially resolve 
the emission line maps and the kinematics of this galaxy. In Section \ref{sec:data}, we introduce the 
observed data. In this section, we also discuss the data reduction procedure and the point spread function (PSF) estimation 
as well as the spectral energy distribution (SED) fitting procedure. The analysis of the IFU data is covered in Section \ref{sec:analysis}. 
The physical properties of the 8 o'clock arc are discussed in Section \ref{sec:integrated}. In Section 
\ref{sec:source}, we introduce our lens modelling technique and also our source reconstruction procedure. 
In this section, we also present the emission line maps and the \hb\ profile in the source plane. The kinematics 
of the galaxy are discussed in Section \ref{sec:dynamic}. We present our conclusions in Section \ref{sec:conclusion}.
\section{Data}
\label{sec:data}
\subsection{NIR spectroscopy with SINFONI}

We obtained $J$, $H$ and $K$ band spectroscopy of the 8 o'clock arc ($\alpha(J2000): 
00^{h} \;22^{m} \;40.91^{s}$ $\delta(J2000): 14^{\circ}\; 31\arcmin \;10.40\arcsec$) using the integral-field spectrograph
SINFONI \citep{eisenhauer03, bonnet04} on the very large telescope in 2009 September (Programme ID: 83.A-0879 A). 
The observation was done in seeing-limited mode with the 0.125 $\rm{arcsec pixel}^{-1}$ scale, for which the total field of view (FOV) 
is 8 arcsec $\times$ 8 arcsec. The total observing time was 4h for $J$, 5h for $H$ and 3.5h for $K$ with individual 
exposure times of 600s. 

\subsubsection{Data Reduction}

The SINFONI data were not reduced with the standard European Southern Observatory pipeline, 
but with a custom set of routines written by N. Nesvadba, which are 
optimized to observe faint emission lines from high-$z$ galaxies. 
These routines are very well tested on SINFONI data cubes for more 
than 100 high-redshift galaxies, and have been used to reduce the 
data presented, e.g., in \cite{lehnert09} and \cite{nesvadba06a,nesvadba06b,nesvadba07a,nesvadba07b}. 

The reduction package uses {\scriptsize {IRAF}} \citep{tody93} standard tools for the
reduction of long-slit spectra, modified to meet the special
requirements of integral-field spectroscopy, and is complemented
by a dedicated set of {\scriptsize {IDL}} routines. Data are dark frame
subtracted and flat-fielded. The position of each slitlet
is measured from a set of standard SINFONI calibration
data which measure the position of an artificial point source.
Rectification along the spectral dimension and wavelength
calibration are done before night sky subtraction to account
for some spectral flexure between the frames. Curvature is
measured and removed using an arc lamp, before shifting
the spectra to an absolute (vacuum) wavelength scale with
reference to the OH lines in the data. To account for the 
variation of sky emission, we masked the source in all frames and 
normalized the sky frames to the average of empty regions in the 
object frame separately for each wavelength
before sky subtraction. We corrected for residuals of the
background subtraction and uncertainties in the flux calibration
by subsequently subtracting the (empty sky) background
separately from each wavelength plane.

The three-dimensional data are then reconstructed and
spatially aligned using the telescope offsets as recorded in
the header within the same sequence of six dithered exposures
(about 1 h of exposure), and by cross-correlating
the line images from the combined data in each sequence,
to eliminate relative offsets between different sequences.
A correction for telluric absorption is applied to each individual 
cube before the cube combination. Flux calibration is carried 
out using standard star observations taken every hour at a 
position and air mass similar to those of the source.

\begin{table*}

\centering                
\begin{tabular}{l c | c | c | c | c | c |}  
\hline\hline    
       &   & A1 & A2 & A3  &  A4 \\
Filter & Band & AB magnitude & AB magnitude & AB magnitude & AB magnitude \\
\hline              
$F450W$  & $B$ &  $21.89 \pm 1.61$ & $21.76 \pm 1.55$ & $21.02 \pm 1.07$ & $22.65 \pm 1.27$\\ 

$F814W$  & $I$ & $ 20.98 \pm 1.03$  & $21.06 \pm 1.07$ & $20.17 \pm 0.7$ &$21.71 \pm 1.44$ \\

$F160W$ & $H$ &$19.35 \pm 0.53$  &  $19.23\pm 0.5$ & $18.46 \pm 0.34$ &$20.35 \pm 0.79$\\
\hline                                
\end{tabular}
 \caption{$HST$ photometry of the 8~o'clock arc images A1-A4. The AB magnitudes correspond to the
total photometry of all components in each image.}
\label{tab:photometry}
\end{table*}
%

\subsection{$HST$ Imaging}

\begin{figure}
\centerline{\hbox{\includegraphics[ width=0.4\textwidth, angle=90]{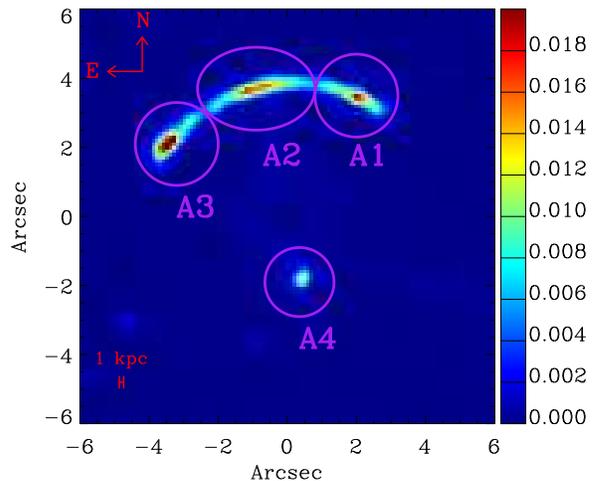}}}
         \caption{The $B$ band $HST$ image (in counts per second) of the 8 o'clock arc is shown. 
         Three images A1-A3 form an arc and A4 is the counter image. The foreground galaxy (lens) 
         has been removed from this image. The scale-bar in this and all following images are at the redshift of the source.}
 \label{fig:8oclock}
\end{figure}

Optical and NIR imaging data of the 8 o'clock 
arc were taken with the Wide Field Planetary Camera 2 (WFPC2) and the
Near Infrared Camera and Multi-Object Spectrometer (NICMOS) 
instruments on the $HST$ (Proposal No. 11167, PI: Sahar Allam). 
The 8 o'clock arc is clearly resolved, and was observed in the five filters WFPC2/$F450W$, 
WFPC2/$F606W$, WFPC2/$F814W$, NIC2/$F110W$, and NIC2/$F160W$, which we will refer to 
as $B$, $V$, $I$, $J$ and $H$ in the following. Total exposure times of 4 $\times$ 1100 s per $BVI$ 
band, 5120 s in the $J$ band, and 4 $\times$ 1280 s in the $H$ band were obtained. 
The $BVI$ frames, with a pixel scale of $0.1$arcsec, were arranged in a four-point dither pattern, 
with random dithered offsets between individual exposures of 1 arcsec in right ascension 
and declination. The $JH$ frames, with a pixel scale of 0.075 arcsec, were also arranged 
in a four-point dither pattern, but with offsets between individual exposures of 2.5 arcsec. 
In order to resolve the 8 o'clock arc better, the $HST$ images were drizzled to obtain 
a pixel scale of 0.05 arcsec. Fig. \ref{fig:8oclock} shows the $B$ band $HST$ image of the 8 o'clock arc and defines the
images A1 through A4 as indicated. 
We performed photometry using the Graphical Astronomy and Image Analysis Tools (GAIA\footnote{http://astro.dur.ac.uk/\%7Epdraper/gaia/gaia.html}). Table \ref{tab:photometry} summarizes the $HST$ photometry of the images A1-A4. 

\begin{figure}
\centerline{\hbox{\includegraphics[trim=3.5cm 0cm 3.2cm 0.5cm, clip=true, width=0.25\textwidth, angle=90]{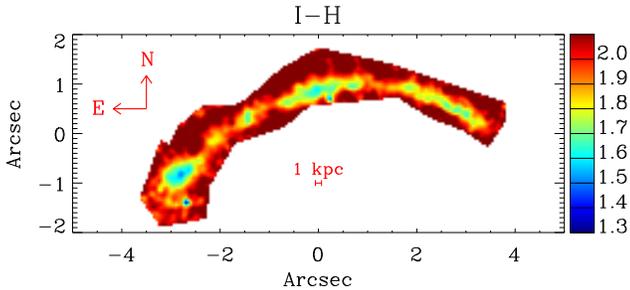}}}
         \caption{The $I-H$ (rest frame $NUV-B$) colour image of the arc is shown. We can see that the substructures of the arc
         are better resolved in this image; for instance, we can resolve better two lensed images of the same clump lying between 
         the A3 and A2 images (see Fig. \ref{fig:reconstruction-450w}, which marks the de-lensed image of this clump with a purple ellipse). 
         We can also resolve two images of another clump which are between the A2 and A1 images (see Fig. \ref{fig:reconstruction-450w}, 
         which notes the de-lensed image of this clump by a red ellipse).}
 \label{fig:8oclock-color}
\end{figure}

As an illustration of the power of the multi-wavelength $HST$ data set, we show the $I$-$H$ (rest frame $NUV-B$) color image of the 
arc in Fig. \ref{fig:8oclock-color}. To construct this we convolved the WFPC2/$F814W$ image to the same PSF as the 
NICMOS/$F160W$ band before creating the color image. 
We can see that the substructures of the arc are better resolved in this image; for instance, we can resolve two individual 
images of the same clump that lie between the A3 and A2 images (see the de-lensed image of the clump 
shown by a purple dashed ellipse in Fig. \ref{fig:reconstruction-450w}).

\subsection{PSF estimation}
We created model PSFs for the $HST$ images using the {\scriptsize{TINY TIM}} package
\footnote{http://tinytim.stsci.edu/cgi-bin/tinytimweb.cgi} \citep{krist11}. 
A measure of the PSF was also obtained using a star in the field. This 
estimate is consistent with {\scriptsize{TINY TIM}} PSFs; however, 
because the star is significantly offset from the arc, in the rest of the paper 
we use only the the {\scriptsize{TINY TIM}} PSFs when analysing the $HST$ data.  

For the SINFONI data we use the standard star observations to estimate the PSF.
The standard star was observed at the end of each observing block at an air 
mass similar to that of the data, in a fairly similar direction, and with the same 
setup. We integrate the standard star cubes in each band along 
the spectral axis to obtain the two-dimensional images of the star. 

We measure the full width at half-maximum (FWHM) size of  the star along the $x$ and $y$ axes of the SINFONI 
FOV by fitting a two-dimensional Gaussian to the resulting image.
We then average the individual measurements of the standard star images 
in each direction to determine the PSF for the corresponding band. The spatial resolutions in right 
ascension and declination are always somewhat different for SINFONI data due to the 
different projected size of a slitlet (0.25 arcsec) and a pixel (0.125 arcsec). The PSFs in the $J$, 
$H$ and $K$ bands are [0.99 arcsec, 0.7 arcsec], [0.8 arcsec, 0.66 arcsec], [0.69 arcsec, 0.51 arcsec], respectively.
%
\section{Analysis of the SINFONI data}
\label{sec:analysis}
\subsection{Nebular emission lines}

The spectrum is first analysed using the \texttt{platefit} pipeline, initially developed for the analysis 
of SDSS spectra \citep[][]{Tremonti04,brinchmann04,brinchmann08} and subsequently modified 
for high-$z$ galaxies \citep[e.g.][] {lamareille06}. 
The nebular emission lines identified in the 8 o'clock arc images A2-A3 are summarized in Table \ref{tab:fluxes} . 
Specifically, the emission lines that we can detect in the spectra of the galaxy are $\oii{\lambda{3727, 3729}}$, 
\hd, \hg, $\heii\lambda{4686}$ and \hb. \hb\ is the strongest detected emission line. In the following, we 
therefore concentrate on this line to further study the dynamical properties of the galaxy.

Due to the redshift of the 8 o'clock arc, we can not study the $\oiii{\lambda4959,5007}$, \ha\ and 
$\nii{\lambda 6548,6584}$ emission lines because they fall outside of the spectral range of the SINFONI bands. 
This means that we can not place strong constraints on the ionization parameter or the metallicity of the galaxy 
using the IFU data.

%
\subsection{The integrated \hb\ profile}
\label{sec:profile}

\begin{figure*}
\centerline{\hbox{\includegraphics[trim=4cm 0.5cm 4cm 0.5cm, clip=true,width=0.45\textwidth, angle=90]
             {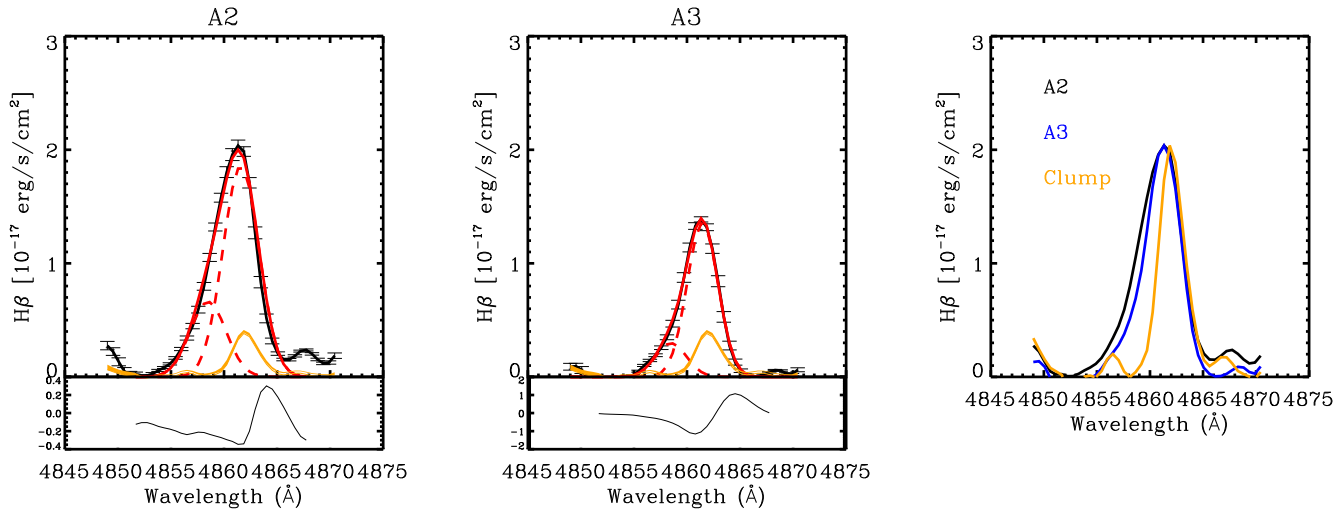}}}
         \caption{Left: integrated \hb\ profile of A2, the highest magnification image, showing two components that we can also resolve in individual spaxels. 
         Middle: integrated \hb\ profile of the A3 image. Both images show the same \hb\ profile (see the right-hand panel). The 
         width of the Gaussian component for both images (red dashed lines) is 1.7 \AA, which gives a velocity dispersion of 104 $\rm{km\; s^{-1}}$. 
         The velocity offset between the two fitted Gaussian components is 278 $\rm{km\; s^{-1}}$ 
         for the A3 image and 191 $\rm{km\; s^{-1}}$ for the A2 image. To illustrate the \hb\ profile of the clump, we add the spectra 
         of two images of the clump and show this profile in orange in both panels (note that the A2 and A3 profiles 
         shown in this figure do not contain the clump profile).
	 The velocity offset between the clump \hb\ profile and the main galaxy \hb\ profile is 126 $\rm{km\; s^{-1}}$. 
	 The bottom panels show the residuals if we fit the \hb\ profile of each image with a single Gaussian. 
	 Right: \hb\ profiles of the A2 and A3 images and the clump are shown; the profiles are normalized to have the same peak. }
 \label{fig:spec-A2-A3}
\end{figure*}

As was noted first by DZ11, the observed Balmer lines of the 8 o'clock arc show asymmetric profiles, this can be seen especially in the \hb line profile. Here we start 
with analysing the integrated \hb\ profile. This offers us, among other things, the possibility of testing our
reduction techniques because in the absence of significant small-scale structure, profiles are expected
to be similar in shape in the different sub-images. We focus here on the spectra of the highest magnification
images, A2 and A3 (see Fig. \ref{fig:8oclock}), and we only integrate over the main galaxy structure, excluding the
clump identified by DZ11. 
The counter image (A4) is complete but is not resolved; the A1 image is only 
partially resolved and is located near the edge of the data cube. The left and middle panel in Fig. \ref{fig:spec-A2-A3}  
show the integrated \hb\ profiles for the images A2 and A3. We can 
see that the two images show the same profile (see the right-hand panel in Fig. \ref{fig:spec-A2-A3}), which is what one expects as they are 
two images of the same galaxy. This result differs from that of DZ11, who 
found different profiles using their long-slit data. They suggested that this might be 
either due to the slit orientation not optimally covering the lensed image A3, or alternatively, 
due to the presence of substructure perturbing the surface brightness of the A2 image. 
Since the IFU data show the same profile for both images,
 we can rule out the possibility that substructure might have caused the differences.
 
We can see that the integrated \hb\ profiles of both images show one main component with 
a broad blue wing; thus, the full profile requires a second Gaussian to be well fitted (see residuals in the bottom panels if we fit one Gaussian to the profiles). \hb\ is a weaker line compared to \ha\ line used basically in the literature to derive the broad and narrow components of the line profiles \citep[e.g.,][]{newman12a,newman12b}. Given our low signal-to-noise (SN) data, a unique broad fit with a physical meaning can not be found considering the fact that residual from sky lines might create broad line widths. For this reason we fit two Gaussian components with the same width to the \hb\ line profile and not a broad and a narrow component. 
We carry out these fits to the \hb\ profiles using the {\scriptsize{MPFIT}} package in {\scriptsize{IDL}}\footnote{http://cow.physics.wisc.edu/$\sim$craigm/idl/mpfittut.html}. During the fitting, we require the lines to have the same velocity widths. These Gaussian components are shown by red dashed lines in the left (A2) and middle panel (A3) of Fig. \ref{fig:spec-A2-A3}.
The width of the Gaussian components for both images is $1.7\pm 0.7$ \AA, which gives a velocity 
dispersion of $104\pm42$ $\rm{km\; s^{-1}}$. The velocity offset between the two fitted Gaussian components which 
are shown by the red dashed curves is $278\pm63.5\; \rm{km\; s^{-1}}$ for the A3 image and 
$191\pm63\;\rm{km\; s^{-1}}$ for the A2 image which are consistent within the errors. 
We can clearly see this blue-shifted component in both images in Fig. \ref{fig:pos-vel-A2} 
and Fig. \ref{fig:pos-vel-A3}. DZ11 fitted two individual Gaussians to the main 
component of the galaxy and concluded that these fits are 
related to the two components (main and clump) with velocity offset of $\sim 61\pm8$ $\rm{km\; s^{-1}}$. 
Since we have resolved the clump using our IFU data and did not include it when integrating the \hb\ profile in
Fig. \ref{fig:spec-A2-A3}, the spectra of the A2 and A3 images plotted 
in Fig. \ref{fig:spec-A2-A3} do not contain any contribution from the clump. To illustrate 
the \hb\  profile of the clump, we add the spectra of the two images of 
the clump and show the total profile with an orange line in three panels in Figure
\ref{fig:spec-A2-A3}. We measure a velocity offset of $126\pm42\; \rm{km\; s^{-1}}$ between the clump and the main 
component of the galaxy. The second component seen for both images (the left Gaussian fits in Fig. \ref{fig:spec-A2-A3}) is 
coming from the part of the arc that was not covered by the slit used by DZ11.
From the lens modelling described in Section \ref{sec:source}, we know that the spatially 
separated blue-shifted component in the A2 image is coming from the north-east part of 
the galaxy (see Fig. \ref{fig:reconstruction-450w}). However, we see from Fig. \ref{fig:pos-vel-A3} that this blue-shifted component of the A3 image is not separated spatially from the main component of the galaxy. The difference between the two images might be due to the fact that the data have insufficient spatial 
resolution to resolve the components in the A3 image. 
%
%

\subsection{Spatially-resolved emission-line properties of the 8 o'clock arc in the image plane}
\label{sec:spatially-resolved}
As we saw above, the integrated \hb\ profile is not well fitted by a single Gaussian, and this is also true for \hg\ and
 can also be seen in individual spatial pixels (spaxels) for \hb. We therefore fit these lines with two or three 
 Gaussian components when necessary. We carry out these fits to the Balmer lines using the {\scriptsize{MPFIT}} package. 
For the same reason explained in the previous section, during the fitting we require the lines to have the same velocity widths. This could be an incorrect approximation in detail but it leads to good fits to the line profiles; the SN and spectral resolution of the data are not sufficient to leave the widths freely variable.
The spatially-resolved \hb\ profiles generally show a main component, which we place 
at a systemic redshift of $2.7363\pm0.0004$ (rest-frame wavelength,$\lambda_{air}=4861.325$) and an additional component that is blue-shifted relative to 
the main component by 120-300 $\rm{km\; s^{-1}}$. The best-fit Gaussian intensity map of these blue-shifted 
and main components of the galaxy are shown in appendix A for the A2 and A3 images. There is also a redshifted component that is detectable close 
to the clumps between the A3 and A2 images (see Fig. \ref{fig:8oclock}). This component is spatially separated from 
the main component by 1 arcsec (mentioned also by DZ11). The velocity difference between 
this component and the main component is $\sim$ 120 $\rm{km\; s^{-1}}$. 

\begin{figure*}
\centerline{\hbox{\includegraphics[width=0.9\textwidth, angle=0]
             {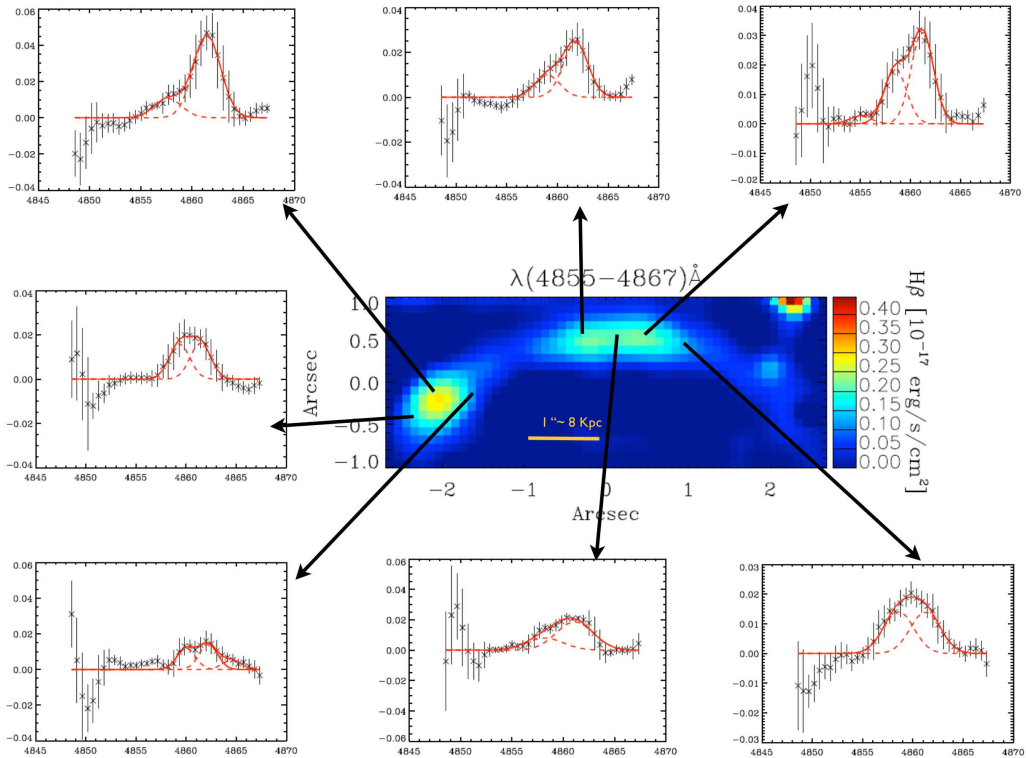}}}
         \caption{The middle panel shows \hb\ line map of the arc. Small panels around the \hb\ line map show the \hb\ profiles of different 
spatial pixels as indicated. The \hb\ line map was integrated over 4855 \AA$<\lambda_{\mathrm{rest}}<$4867 \AA. We can see that \hb\ shows different profiles 
at different pixels, which are composed of multiple components.}
 \label{fig:Hb-profile}
\end{figure*}

The central map in Fig. \ref{fig:Hb-profile} shows the spatial distribution of \hb\ line flux across the main components of the arc,
where we have integrated the line flux between $\lambda=4855$\AA\ and 4867\AA. The small 
panels around the \hb\ line map show the \hb\ profiles of different spatial pixels as indicated.  These individual panels
clearly show that the  \hb\ line shows different profiles in different regions across the lensed images. 

\begin{figure}
\centerline{\hbox{\includegraphics[width=0.5\textwidth, angle=90]
             {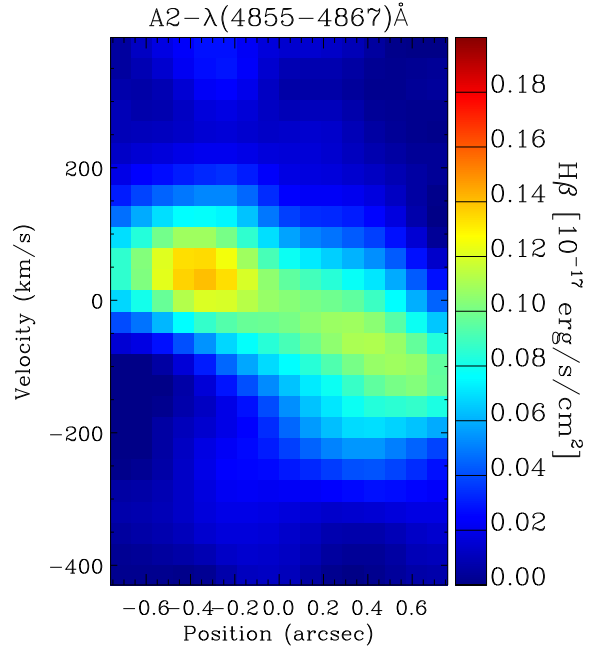}}}
         \caption{Position-velocity diagram of the A2 image, with negative values corresponding to blue shift. There are two components that are clearly offset both spatially and in the velocity direction. The velocity offset between the two components is $\sim$ 130 $\rm{km\; s^{-1}}$, and the spatial separation between them is $\sim$ 1 arcsec. Position is relative to the center 
	of the A2 image along the length of the arc.}
 \label{fig:pos-vel-A2}
\end{figure}

\begin{figure}
\centerline{\hbox{\includegraphics[width=0.5\textwidth, angle=90]
             {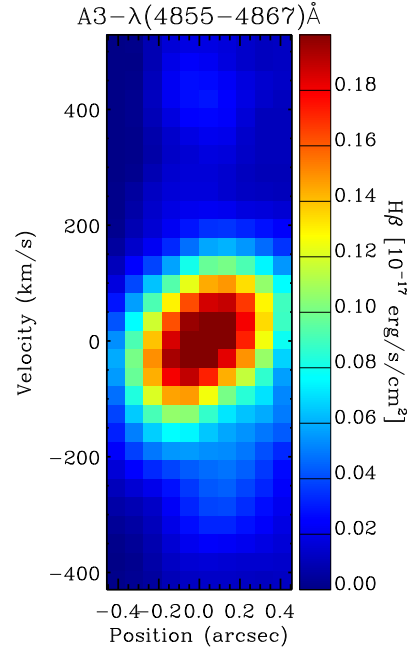}}}
         \caption{Position-velocity diagram of the A3 image. This image shows a blue-shifted component
         that is not spatially separated, in contrast to Fig.~\ref{fig:pos-vel-A2}. Position is relative to the center of the A3 image along the length of the arc.}
 \label{fig:pos-vel-A3}
\end{figure}

We can show these components in an alternative way, using the position-velocity diagrams in Fig. \ref{fig:pos-vel-A2} 
and Fig. \ref{fig:pos-vel-A3} for the A2 and A3 images, respectively. Fig. \ref{fig:pos-vel-A2} clearly 
shows two spatially separated components corresponding to the A2 image. The peak of one component is blue-shifted by 
$\sim$130 $\rm{km\; s^{-1}}$ and spatially separated by $\sim\; 1\arcsec$ relative to the peak of the other. DZ11 identified 
these two components with the main galaxy and the clump because 
they could not separate the clump from the rest of the galaxy using long-slit observations. Here, 
using IFU data, we have excluded the clump from these position-velocity diagrams. The two retained components are associated with the 
galaxy and the red (in the spectral direction) component that DZ11 identified as the 
clump is part of the main galaxy. From the lens modelling that we describe in Section \ref{sec:source}, 
we will see that the blue-shifted component comes from the eastern part of the galaxy (see Fig. \ref{fig:reconstruction-450w}). 
The A3 image in Fig. \ref{fig:pos-vel-A3} also shows this blue-shifted component but not as spatially separated. 
We will argue below (see Section \ref{sec:outflow}) that a reasonable interpretation
of this component might be that it corresponds to an outflow from the galaxy.

%
%
\section{The physical properties of the 8 o'clock arc}
\label{sec:integrated}

\subsection{SED fitting}

\begin{figure}
\centerline{\hbox{\includegraphics[width=0.37\textwidth, angle=90]
             {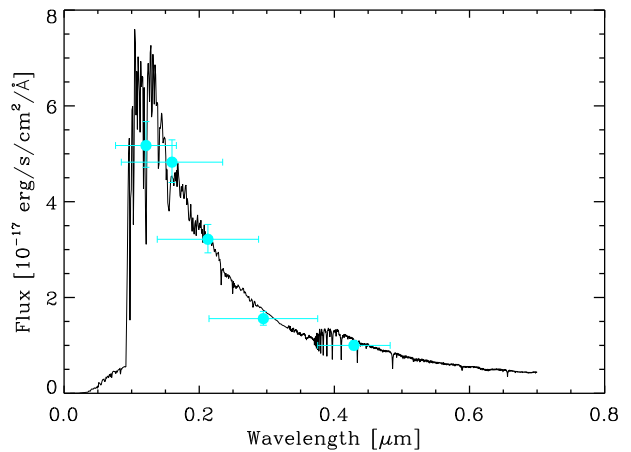}}}
         \caption{$HST$ photometry for the 8 o'clock arc in the rest-frame (cyan filled circles) with the best-fit SED (solid curve) over plotted. 
         The horizontal error bars show the wavelength coverage of the $HST$ filters.}
 \label{fig1}
\end{figure}

To determine the physical parameters of the 8 o'clock arc, we fit 
a large grid of stochastic models to the $HST$ $BVIJH$ 
photometry to constrain the SED. 
The grid contains pre-calculated spectra for a set of 100,000 different 
star formation histories (SFH) using the \citet[BC03]{bc03} population synthesis 
models, following the precepts of \citet{gallazzi05,gallazzi08}. Fig. \ref{fig1} 
shows the best-fit SED. We corrected the observed magnitudes for galactic 
reddening. We corrected the photometry for Galactic foreground dust extinction using $E(B -V)_{Gal} = 0.056$
 \citep{schlegel98}. 

We follow the Bayesian approach presented by \cite{kauffmann03} to
calculate the likelihood of the physical parameters. We take 
the median values of the probability distribution functions (PDFs) as our best 
estimated values. In particular, the parameters we extract are 
the stellar mass, $\rm{M_{\star}}$,  the current star-formation rate, $\rm{SFR_{SED}}$, 
the dust attenuation, $\tau_{V}$ and the $r$-band luminosity weighted age. The physical 
parameters from the SED fitting are summarized in Table \ref{tab:sed-param}.

DZ11 also carried out SED fitting to the photometric data for the 8 o'clock arc. They explored 
the cases with and without nebular emission \citep{schaerer09, schaerer10}. Since their results do 
not change significantly, we do not consider the effect of nebular emission in our study. 
They also included photometry from IRAC, which in principle should improve 
constraints on stellar masses. We have opted against using these data, keeping the higher spatial resolution of $HST$ + SINFONI,
as the stellar mass from our fits is only slightly higher, but consistent with their results within the errors, and this
is the quantity of most interest to this paper.

\begin{figure}
\centerline{\hbox{\includegraphics[width=0.37\textwidth, angle=90]
             {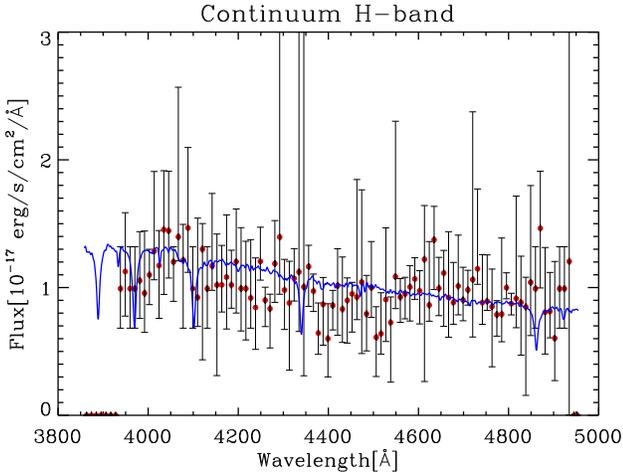}}}
         \caption{Median continuum in the $H$-band calculated in bins of 10\AA\ (red filled circles). Error bars 
         show 16th and 84th percentiles of the distribution around the median. Estimated model continuum from SED fitting is shown by the blue solid curve.}
 \label{fig:continuum-Hband}
\end{figure}

To compare our SED fit to the observed continuum spectrum, we estimate the continuum in 
the SINFONI spectra by taking the median of the spectrum in 
bins of 10 \AA. Fig. \ref{fig:continuum-Hband} shows the $H$-band median continuum, summed 
over all images in the arc, in comparison to the estimated model continuum from the SED fitting. The agreement is satisfactory,
although the SN of the continuum precludes a detailed comparison.
%

%
%
\begin{table*}
\begin{center}
\begin{tabular}{ccccccc}
 \hline
 \multicolumn{1}{l}{Image} &
\multicolumn{1}{l}{$\log\; (M_{\star}/M_{\odot})$}  & \multicolumn{1}{l}{$\log (SFR/{M_{\odot}\;yr^{-1})}$ } &
\multicolumn{1}{l}{$\log (sSFR/{yr^{-1})}$} &   \multicolumn{1}{l}{$\log (age/{yr)}$} & 
\multicolumn{1}{l}{$\tau_{V}$} & \multicolumn{1}{l}{\rm{$\log (Z/Z_{\sun})$}}\\
\hline
\hline
\\
A2 & $10.24^{9.99}_{10.69}$ & $1.86^{1.71}_{1.96}$ & $-8.47^{-8.98}_{-8.17}$ & $8.32^{7.68}_{8.93}$ & $0.17^{0.07}_{0.3}$ & $0.08^{-0.19}_{0.2}$\\
\hline
\\
A3 & $10.32^{10.08}_{10.77}$ & $1.96^{1.82}_{2.07}$ & $-8.44^{-8.93}_{-8.15}$ & $8.28^{7.67}_{8.90}$ & $0.17^{0.07}_{0.3}$ & $0.07^{-0.2}_{0.2}$\\
\hline
\end{tabular}
\end{center}
\caption{Physical parameters derived from SED fitting. Parameters from left to right are stellar mass, 
$\rm{M_{\star}}$, current star-formation rate, SFR, specific star formation rate (sSFR), $r$-band luminosity weighted age, dust attenuation 
$\tau_{V}$ and metallicity.}
\label{tab:sed-param}
\end{table*}

\subsection{Parameters derived from emission line modelling}
\label{sec:derived}

To derive physical parameters for the ionized gas in the 8 o'clock arc, we make use of a grid of  \citet[][hereafter CL01]{cl01} models. 
We adopt a constant SFH and adopt the same grid
used by \citet[][hereafter B04; see appendix A in \cite{shirazi12} and B04 for further details]{brinchmann04}. 
In total, the model grid used for the fits has $2\times10^5$ different models. 
The model grids and corresponding model parameters are summarized in Table \ref{tab:summary2}.  
Our goal here is to derive representative overall parameters for the galaxy, and since the fitting methodology outlined in B04 
works best with $\oii{\lambda3727}$, $\hb$,
$\oiii{\lambda4959}$, $\ha$, and $\nii{\lambda6584}$ all available, 
we here take the emission line measurements from DZ11 since  the last three lines fall outside the spectral range of our SINFONI data cube.
For the quantities that only depend on line ratios, i.e., all but the SFRs, this is appropriate, but for the 
SFR we need to correct for light missed by the long slit observations of DZ11, and we do this by normalizing 
to the \hb\ line flux from the SINFONI data.

We use the same Bayesian methodology as for the SED fit and again take the median value of each PDF to be the best estimate
of a given parameter and the associated $\pm 1 \;\sigma$ confidence
interval to be spanned by the 16th--84th percentiles of the PDF.

\begin{figure*}
\centerline{\hbox{\includegraphics[width=0.75\textwidth, angle=90]
             {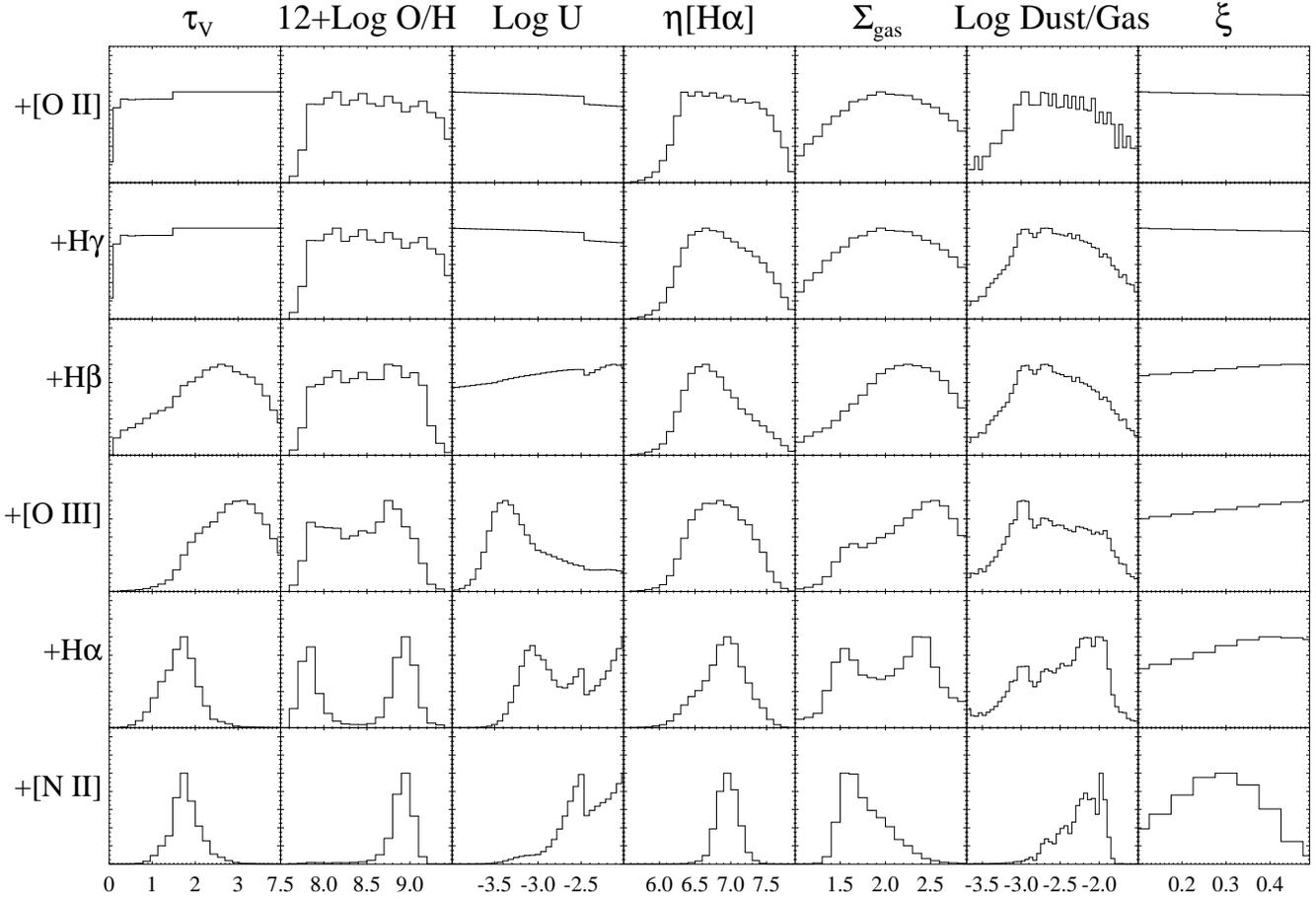}}}
         \caption{The PDFs for optical depth in the $V$-band, 
         the gas phase oxygen abundance,  the logarithm of the ionization parameter,
       $\log_{10}$  of the conversion factor from observed \ha\ luminosity to star formation
       rate, the logarithm of the gas surface density in $M_{\odot}/\mathrm{pc}^2$,  the
       log of the DGR and the metal-to-gas ratio. From one row to the next we include the 
       line indicated on the left side in the fitting in addition to the previous emission lines. For the observational data in this figure we make use of
       emission line fluxes measured by DZ11, from their Table 2, for image A3 (see Fig. \ref{fig:8oclock}).
	Our best-fit parameters and the associated 16th-84th percentiles are summarized in Table \ref{tab:cl01}. 
	The abundance of oxygen reported by DZ11 is lower than what we find. }
 \label{fig:pdf}
\end{figure*}

In Fig. \ref{fig:pdf} we illustrate our technique by showing the
effect on the PDFs of parameters, when we fit a model
to an increasing number of the emission lines. We start with $\oii{\lambda3727}$ and show how we
get more well defined PDFs for the indicated parameters as we add the
emission lines indicated on the left.  We show the PDFs
for the dust optical depth in the $V$-band, the gas phase oxygen
abundance, the ionization parameter, the conversion factor from \ha\ luminosity to SFR (see CL01 for further details), the gas mass surface 
density, the dust-to-gas ratio (DGR) and the metal-to-dust ratio of the ionized gas. The latter three
quantities are discussed in some detail in \citet[][hereafter B13]{brinchmann13} and we 
discuss them in more detail below. The resulting PDFs are shown for the A3 image and 
the best-fit parameters derived from the final PDFs are summarized in Table \ref{tab:cl01}.

We note that high electron density values ($n_e>100\; \rm{cm^{-3}}$) are not included in the CL01 models. The only parameter which is density dependent in Fig. \ref{fig:pdf} is the ionization parameter which as you see in the figure is not constrained by the model.

The oxygen abundance reported by DZ10 and DZ11 is lower than what we find ($8.93_{-0.07}^{+0.09}$), DZ11 find $8.46\pm0.19$ for the A3 image; however, gas metallicity derived by \citep[][$8.58\pm0.18$]{finkelstein09} is more consistent with our results. 
The lower metallicity that they derive is not entirely surprising for two reasons. First, it is well-known \cite[e.g.][]{kewley08} 
that metal abundance estimators show significant offsets, so even when converted to a 
solar scale, one has to accept a systematic uncertainty in any comparisons that use different methods 
for metallicity estimates. Secondly, the estimates in DZ10 and DZ11 primarily rely on the calibration relationships 
(N2 calibration) from \cite{pettini04}, which are based on local \hii\ regions and an extrapolation to higher metallicity. 
The use of these calibrated relationships implicitly assumes that the relationship between ionization parameter 
and metallicity is the same at high and low redshifts. This is a questionable assumption; indeed the electron density 
we find for the 8 o'clock arc is considerably higher than seen on similar scales in similar galaxies at low redshift (see 
Fig. \ref{fig:eoc-vs-local}), indicating that the U-Z relationship is different at high redshift, and thus that the N2 
calibration is problematic. In our modelling we leave U and Z as free variables; thus, we are not limited by this. It is 
difficult to ascertain which approach is better but the advantage of our approach for the 8 o'clock arc is that it 
uses exactly the same models which are used to fit local SDSS samples.

We note that our gas metallicity estimates are more consistent with stellar and the ISM metallicity derived by DZ10. Considering a 0.2 dex systematic uncertainty, our results are also consistent with DZ11 gas metallicity.

It is well-known that the estimation of ISM parameters from strong emission lines is 
subject to systematic uncertainties (see however B13 for an 
updated discussion). To reduce the effect of these uncertainties, we have also assembled
a comparison sample of star-forming galaxies at $z\sim 0.1$ from the SDSS. We used
the MPA-JHU value added catalogues \citep[B04;][]{Tremonti04} for SDSS DR7\footnote{\url{http://www.mpa-garching.mpg.de/SDSS/DR7}} 
as our parent sample. We define a star-forming 
galaxy sample on the basis of the \nii{6584}/\ha\ versus \oiii{5007}/\hb\ diagnostic diagram, often 
referred to as the BPT diagram \citep{BPT}. For this we used the procedure detailed in B04 with 
the adjustments of the line flux uncertainties given in B13. From this parent sample, we select
all galaxies that have stellar mass within 0.3 dex of the value determined for the 8 o'clock arc and whose SFR
is within 0.5 dex of the 8 o'clock arc, based on the parameters determined from SED fitting to the A2 image 
(Table~\ref{tab:sed-param}). This resulted in a final sample of 329 galaxies, which we 
compare to the 8 o'clock arc below. 


\begin{table}
\begin{center}     
\centering                
\begin{threeparttable}   
\begin{tabular}{l c | c  c c }
\hline 
Line & $\lambda_{air}$\,(\AA) & \multicolumn{1}{c|}{A2} & \multicolumn{1}{c}{A3} & \\
\hline\hline                

$\oii$                           & 3726.032     &  $65.8\pm 1.9$ &  $36.2\pm 1.6$  \\
$\oii$                          & 3728.815     &  $58.2\pm 1.8$ &  $29.8\pm 1.5$  \\ 
$\hd	$                          & 4101.734   &   $25.6\pm 0.6$ &   $12.9\pm 0.3$  \\
$\hg$                        & 4340.464    &  $45.3\pm 0.6$ &  $26.3\pm 0.3$  \\
$\hb$                        & 4861.325     &  $102.6\pm 1.4$ &   $75.8\pm 1$  & \\
\hline                               
\end{tabular}

  \end{threeparttable}
\end{center}
\caption{Nebular emission lines identified in the 8~o'clock arc images A2 and 
A3 and their fluxes. Fluxes are not corrected for lens magnification.
Integrated line fluxes are in units of $10^{-17} \rm{erg \;s^{-1} cm^{-2}}$.}  
\label{tab:fluxes} 
\end{table}
%
%
\begin{table}
\begin{center}
\begin{tabular}{|l|l|}
 \hline
\multicolumn{1}{l}{Parameter}  &
\multicolumn{1}{l}{Range} \\
\hline
\hline
$Z$,  metallicity & $-1<\rm{\log(Z/Z}_{\odot})<0.6$, 24 steps \\
$U$,  ionization parameter & $-4.0<\rm{\log U}< -2.0$, 33 steps \\
$\tau_{V}$, total dust attenuation & $0.01<\tau_{V}<4.0$, 24 steps \\
$\xi$, dust-to-metal ratio & $0.1<\xi<0.5$, 9 steps \\
\hline
\end{tabular}
\end{center}
\caption{The CL01 model grid used in the present work.}
\label{tab:summary2} 
\end{table}
%
%
\begin{table}
\begin{center}
\begin{threeparttable}
\begin{tabular}{lccc}
 \hline
 & 16th\tnote{a} & median \tnote{b} & 84th \tnote{c}\\
\hline
\hline

$\log (Z/Z_{\odot})$\tnote{d} &   0.046 & 0.130 &  0.196  \\ 
$\log$ U\tnote{e}                       &   -2.6 & -2.2 & \ldots\tnote{f}  \\ 
$\tau_{V}$\tnote{g}                &   1.1 &  1.6 & 1.8 \\
SFR $\,{(M_{\odot}\;yr^{-1})}$\tnote{h}           &  157 & 165 & 173 \\ 
$\log (\Sigma_{\rm{gas}}/{M_{\odot}\;pc^{-2})}$\tnote{i} & 1.46 & 1.60 & 1.87 \\
12+Log O/H &    8.86 & 8.93 & 9.02\\
\hline 
\end{tabular}

\begin{tablenotes}
       \item[a] The 16$^{\mathrm{th}}$ percentile of the PDF of the given quantity.
        \item[b] The 50$^{\mathrm{th}}$ percentile, or median, of the PDF of the given quantity.
        \item[c] The 84$^{\mathrm{th}}$ percentile of the PDF of the given quantity.
        \item[d] The log of the total gas-phase metallicity relative to solar.
        \item[e] The log of the ionization parameter evaluated at the edge of the Str{\"o}mgren sphere (see CL01 for details).
        \item[f] The electron density in the 8 o'clock arc is higher than that assumed in the CL01 models, and the ionization parameter is therefore close to the edge of the model grid, to which we do not quote an upper limit.
        \item[g] The dust attenuation in the $V$-band assuming an attenuation law $\tau(\lambda) \propto \lambda^{-1.3}$.
        \item[h] The SFR.
        \item[i] The log of the total gas mass surface density.
         \end{tablenotes}

\end{threeparttable}
\end{center}
\caption{Physical parameters of the ISM derived from the spectrum of the A3 image.}
\label{tab:cl01}
\end{table}
%
%
%

\subsection{AGN contribution}

The preceding modelling assumes that the ionizing radiation in the 8 o'clock arc is dominated by stellar sources.  The position of the 8 o'clock arc in the BPT diagnostic diagram \citep[see Fig. 6 in][]{finkelstein09}, 
which has been widely used for classifying galaxies, does, however, suggest that the emission lines from this galaxy 
might be contaminated by an AGN. However, there is some evidence indicating that the AGN contribution for this galaxy is negligible.
 
First, high-resolution Very Large Array (VLA) imaging at 1.4 and 5 GHz show that, although there is a radio-loud AGN 
associated with the lensing galaxy and the arc is partially covered by the radio-jet from this AGN, there is no 
detectable radio emission from the unobscured region of the arc down to a $3\sigma$ flux-density limit of 108 
$\mu \rm{Jy\; beam^{-1}}$ \citep{volino10}.  
Secondly, we can detect $\heii{\lambda 4686}$ for this galaxy, a high-ionization line that is very sensitive to 
the AGN contribution. Therefore, we can use this line as a probe 
to estimate the AGN contribution to the spectrum of this galaxy. We use a new diagnostic diagram of $\heii/ \hb$ 
versus $\nii/\ha$ introduced by \cite{shirazi12} to calculate this. 
As we do not have $\nii$ and \ha\ from the SINFONI observation, we use the DZ11 estimates for these emission 
lines. The $\heii{\lambda 4686}/\hb$ is not very sensitive 
to electron temperature and metallicity. Therefore using the DZ11 estimate for $\nii/\ha$ is sufficient 
for us to locate the position of this galaxy in the $\heii/\hb$ diagram. 
\cite{shirazi12} derive an almost constant line versus metallicity at which the contribution of an AGN to the $\heii$ 
emission amounts to about 10\%. They showed if the $\heii$ is contaminated 
by this amount, the total AGN contribution to other emission lines in the spectrum of the galaxy is less than 1\% 
\cite[see fig. 3 in][]{shirazi12}. As the position of the 8 o'clock arc in this diagram ($\log \; \heii/\hb=-1.4$) is below the above mentioned line, we can conclude 
that the contribution of AGN to the optical emissions is negligible. 

We note that the broad $\heii{\lambda1640}$ emission found by DZ10 can be affiliated to the presence
of Wolf-Rayet stars \citep[see also][]{eldridge12}.


\subsection{SFR and dust extinction}

We have two main methods available to determine the SFR of the 8 o'clock arc from its 
emission line properties.  We can use the SFR derived from the emission line fits described above, but to provide spatially resolved
SFR maps we need to turn to the lines available in the SINFONI data cube. The  $\ha {\lambda 6563}$ emission line is commonly
 used as a SFR indicator at 
low redshift \citep{kennicutt98}. Unfortunately, for the redshift of the
8 o'clock arc, \ha\ falls outside of the spectral range of the $K$ band of SINFONI, and we can not use this indicator to derive 
the spatially resolved SFR. We are therefore limited to using $\hb{\lambda 4861}$ as a  tracer of the spatially resolved SFR.
The advantages and disadvantages of 
using this indicator to measure the SFR were originally discussed by \cite{kennicutt92} and were studied in detail by 
\citet{moustakas06}. In comparison to \ha, \hb\ is more affected by interstellar dust and is more sensitive to the underlying 
stellar absorption \citep[see section 3.3 and fig. 7 in ][]{moustakas06}.

We use the empirical SFR calibrations from \citet[][Table 1]{moustakas06}, parametrized 
in terms of the $B$-band luminosity, to calculate the SFR from \hb\ luminosity:
\begin{equation}
SFR\;({M_{\sun} yr^{-1}}) =10^{-0.24}  \times10^{0.943} \times10^{-42}\frac{\rm{L({H_\beta} )}}{erg\; s^{-1}}.
\label{eq:one}
\end{equation}
\citet{moustakas06} derived the SFR calibration assuming a Salpeter initial mass function \citep[IMF,][]{salpeter55} over 0.1-100 ${M}_{\sun}$. 
The correction factor of $10^{-0.24}$ in Equation \eqref{eq:one} is used to correct to a Chabrier IMF \citep{chabrier03}.
We interpolate between bins of $\rm{L(B)}$ to obtain the relevant conversion factor from the \hb\ luminosity to SFR.

We use a dust extinction $E(B-V)=0.3\pm0.1$ derived from the $\hg/\hb$ Balmer line ratio with an intrinsic 
$\hg/\hb=0.468$ to correct both SFRs for dust extinction. This is consistent with the estimate 
of DZ11 within the errors. We used magnification factors of $\mu^{\rm A2} = 6.3$ and 
$\mu^{A3} = 4.9$ to correct the SFR estimates for the effect of gravitational lensing. 
The magnification calculated using the lens modelling is described in Section {\ref{sec:source}. 
We use the 0.012 contour level (in count per second unit) in the $B$ $HST$ image for detecting individual images.

We measure $\hb= (102.6 \pm 1.4) \times 10^{-17}\; \rm{erg\; s^{-1} cm^{-2} }$ for the A2 image 
corresponding to an observed SFR of $228 \pm 10.9$ ${M_{\sun} \;yr^{-1}}$ and 
$\hb= (75.8 \pm 1.) \times 10^{-17}\; \rm{erg\; s^{-1} cm^{-2} }$ for the A3 image, corresponding 
to an observed SFR of $227 \pm 10.5$ ${M_{\sun} \;yr^{-1}}$(corrected for gravitational lensing 
magnification and dust extinction).

We can contrast this result to the integrated SFRs derived for the A2 and A3 images by fitting the CL01 models, 
after scaling the DZ11 line fluxes to match the \hb\ flux from the SINFONI cube. These are
$160\pm 12$ ${M_{\sun} \;yr^{-1}}$ and $165 \pm 10.5$ ${M_{\sun} \;yr^{-1}}$, respectively. These values are somewhat discrepant but we
note that systematic uncertainties are not taken into account in the calculation here. 
 
More importantly, the \hb\ calibration allows us to calculate maps of the spatial distribution of the optically visible
 star formation in the 8 o'clock arc. Furthermore, we can make use of our decomposition of the \hb\ profile to 
 calculate maps for each component. This is shown in Fig. \ref{fig:bgr} which shows these three calculated SFR maps for the A2 image. These were derived by integrating over 
the blue ($\lambda(4855-4859)$\AA), green ($\lambda(4859-4863)$\AA), and red ($\lambda(4863-4867)$\AA) 
parts of the \hb\ profile. We can see that the blue and red maps peak at
different part of the image, which suggests that they represent different components of the galaxy. 

\begin{figure}
\centerline{\hbox{\includegraphics[width=0.5\textwidth, angle=0]
             {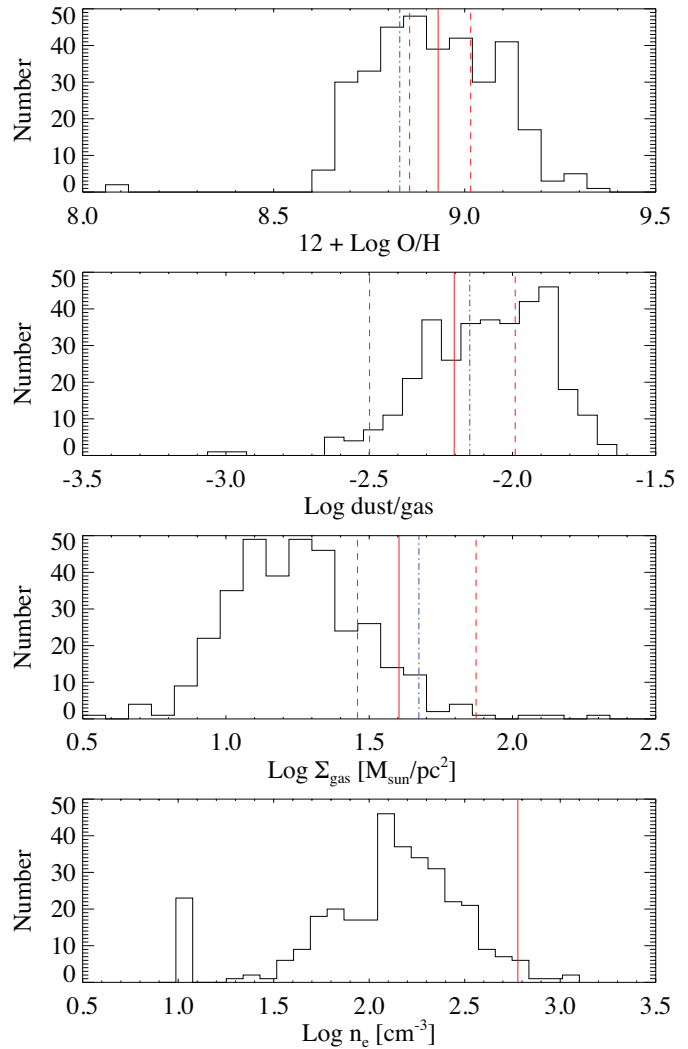}}}
\caption{Top panel: distribution of oxygen abundance for the local
	comparison sample. The solid red line shows the median value derived for image
	 A3 with the dashed vertical lines showing the 16\% and 84\% percentiles. The
	  dash-dotted line shows the median estimate for image A2 and is consistent with that
	  derived for A3. Second panel: same for the DGR. Third panel: 
	  same for gas surface mass density. Bottom panel:
	  distribution of electron densities in the comparison sample. These were derived using the ratio of the \sii{6716,6731} lines assuming an electron temperature of $T_e=10^4\;\mathrm{K}$. The estimated electron density for the 8 o'clock arc, derived from the ratio of the \oii{3726,3729} lines, is shown by the solid red line.}
\label{fig:eoc-vs-local}
\end{figure}

\subsection{Metallicity and DGR}

The modelling described above gives an estimate of the oxygen abundance and the DGR of the 8 o'clock arc. Our best-fit oxygen abundance, $12 + \log \mathrm{O/H}=8.93^{+0.09}_{-0.07}$ (random errors only), is consistent with the determination by DZ11 within the error. In the top panel of Fig. \ref{fig:eoc-vs-local} we compare this value to that determined for our local comparison sample. The values for A3 are shown as the solid red lines, with the 1$\sigma$ confidence interval indicated by the dashed red lines. We also calculate values from the fluxes provided for image A2, and these are shown as the dot-dashed blue lines. We suppress the uncertainty estimate for the latter estimate but it is similar to that of A3; thus, the two measurements are consistent given the error bars. 

Two points are immediately noticeable from this plot. First, the oxygen abundance is mildly super-solar, (the solar oxygen abundance in the CL01 models is 8.82), and secondly, the value for the 8 o'clock arc is the same as for the local comparison sample. Since the local sample was selected to have approximately the same stellar mass and SFR as the 8 o'clock arc, we must conclude that the 8 o'clock arc lies on the stellar mass--oxygen abundance--SFR manifold found locally \citep{mannucci10}. A similar conclusion was also found by DZ11.

The second panel in Fig. \ref{fig:eoc-vs-local} shows the DGR of the SDSS comparison sample as a histogram and the values for the 8 o'clock arc as the vertical lines as in the top panel. The uncertainty in $\log \mathrm{DGR}$ is fairly large, but there is no evidence that the 8 o'clock arc differs significantly from similar galaxies locally. As we will see next, this does not necessarily imply that the ISM has all the same properties.

\subsection{The gas surface mass density}
\label{sec:gas}
It is of great interest to try to estimate the gas content of high redshift galaxies in general, and we have two methods to do
this for the 8 o'clock arc. The first is to use the Kennicutt--Schmidt relation \citep{schmidt59,kennicutt98} between the gas surface density 
and SFR per unit area to convert our spatially resolved SFRs to gas surface densities.  
We can then use this estimated gas surface density to calculate the gas mass. The estimated gas surface density is plotted in Fig. \ref{fig:gas-surf} and is simply a transformation of the SFR map shown earlier. We can then integrate this surface density of gas to get an estimate of the 
total gas mass. With the canonical calibration of Kennicutt (1998) this gives an estimate of $\log (M_{\mathrm{gas}}/M_{\sun})= 10.43 \pm 1.18$. If instead we use the calibration used by DZ11  \citep{bouche07}, we find a total gas mass of $10.30\pm1.20$. In either case, the gas content is comparable to the total stellar mass of the galaxy. 

We can also estimate the gas content in a way independent of the SFR by exploiting a new technique presented in B13, which exploits the temperature sensitivity of emission lines to provide constraints on the dust-to-metal ratio; together with an estimate of the dust optical depth (see B13 for detail), one can derive a constraint on the surface mass density of gas (molecular plus atomic). B13 show that this technique
can give gas surface densities in agreement with H \textsc{i}+H$_2$ maps to within a factor of 2. The results of our fits are shown in Fig.~\ref{fig:pdf}. In passing, we note that this technique, in contrast to the Kennicutt-Schmidt method, only relies on line ratios and these lines all originate in much the same regions; thus, lensing amplification should be unimportant here.

The median estimate is $\log (\Sigma_{\rm{gas}}/{M_{\odot}\;pc^{-2})}= 1.60$, which is consistent with the value that was derived using the Kennicutt-Schmidt relation but derived
in an independent manner, and crucially using a method that is formally independent of the SFR. We contrast the estimate of $\log \Sigma_{\mathrm{gas}}$ for the 8 o'clock arc to the 
SDSS comparison sample in the third panel of Fig. \ref{fig:eoc-vs-local}. The median value for the 
local sample is $\Sigma_{\mathrm{gas}} \approx 17\;M_{\odot}\;\mathrm{pc}^{-2}$, while the median estimate for the 8 o'clock arc is $\Sigma_{\mathrm{gas}} \approx 40\;M_{\odot}\;\mathrm{pc}^{-2}$.  Since 
this comparison is differential and does not depend on scaling relations calibrated on local
samples, the conclusion that the surface density of gas in the 8 o'clock arc is more than
twice that of similar $z\sim 0.1$ galaxies should be fairly robust. Thus, while the 8 o'clock
arc does lie on the $M_*$--$\mathrm{O/H}$--SFR relation, it has a significantly higher
gas surface density than galaxies lying on the same relation locally. 
This highlights the fact that even though a particular scaling relation is established at
high-$z$, it does not imply that galaxies lying on the relation are necessarily similar
to local galaxies.


\begin{figure} 
\centerline{\hbox{\includegraphics[width=0.67\textwidth, angle=90]
             {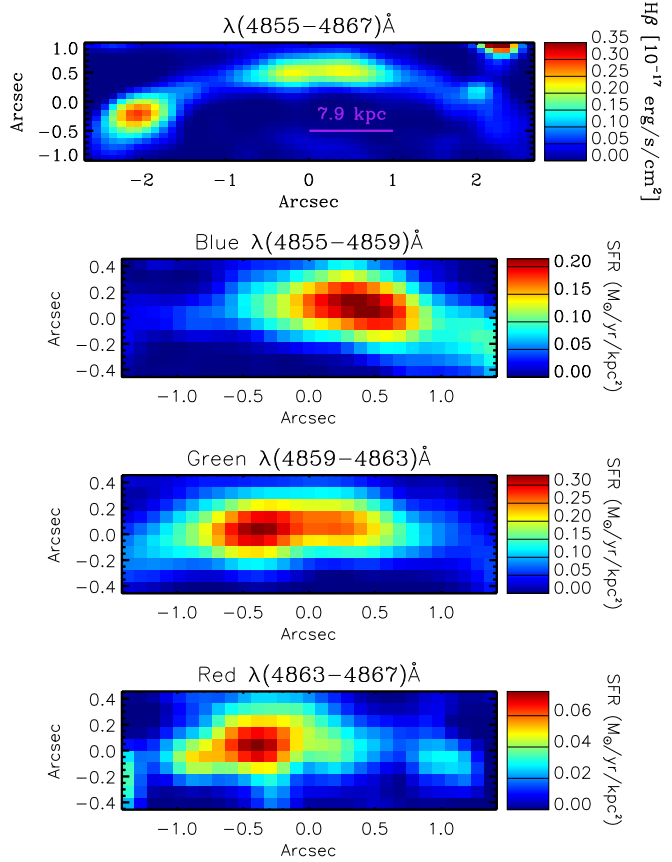}}}
         \caption{Top: \hb\ line map of the arc. This map is not corrected for magnification. 
         Lower panels: three different calculated SFR (corrected for lensing magnification and dust extinction) 
         maps for the A2 (middle) image. SFRs are derived by integrating over the blue, green and red parts of the \hb\ profile, respectively from top to bottom. 
         The wavelength ranges covered are indicated above the SFR maps. The colour scale 
         shows the SFR per $\rm{kpc}^2$ as indicated.}
 \label{fig:bgr}
\end{figure}
\begin{figure}
\centerline{\hbox{\includegraphics[trim=3.5cm 0cm 3cm 0cm, clip=true,width=0.25\textwidth, angle=90]
             {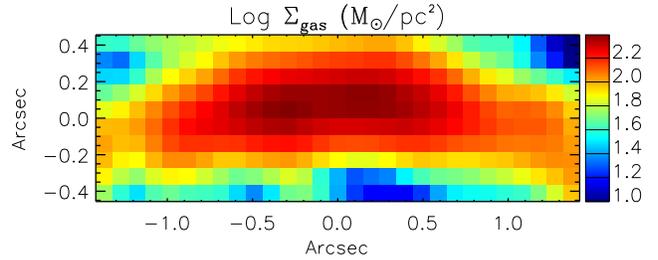}}}
         \caption{The resolved gas surface density for the A2 image estimated using the Kennicutt-Schmidt relation. }
 \label{fig:gas-surf}
\end{figure}



\subsection{Electron density}
\label{sec:density}

We can estimate the electron density using the $\oii{\lambda 3729}/\oii{\lambda 3726}$ ratio. 
Both the X-shooter and the SINFONI observations of the 8 o'clock arc give similar ratios for the 
$\oii{\lambda 3729, 3726}$ lines. The SINFONI observations give a value of 
$\oii{\lambda 3729}/\oii{\lambda 3726} =0.88$ which corresponds to a high electron density of 
$n_e \sim 600\; \rm{cm}^{-3}$ for this galaxy. High electron densities ($n_e >250\; \rm{cm}^{-3}$)
have been found in many other high-$z$ galaxies \citep[see e.g.][]{lehnert09, lehnert13, wuyts12, shirazi13}. \citet{shirazi13} show that high-$z$ galaxies that they study have a median value of eight times higher electron densities compared to that of their low-$z$ analogs (median $n_e$ for local galaxies is $n_e \sim100 \;\rm{cm}^{-3}$).
For the local comparison
sample we are unable to reliably estimate $n_e$ from the \oii{} line ratio, so we use the 
\sii{6717,6731} lines instead. These values are compared to that for the 8 o'clock arc in
the bottom panel of Fig. \ref{fig:eoc-vs-local} and as that figure makes clear, the 8 o'clock arc has
significantly higher electron density than similar SDSS galaxies. Note that the \sii{} ratio is not very
sensitive to electron density variations at $n_e<100\; \rm{cm}^{-3}$, hence the somewhat truncated shape of the distribution there.

The high electron density is likely to lead to a high ionization parameter (see Fig.~\ref{fig:pdf}). Indeed, the 8 o'clock arc lies close to the maximum ionization parameter
in the CL01 models, and its electron density is well above the value assumed
 ($n_e=100\;\mathrm{cm}^{-3}$) in the CL01 model calculations. For this reason, we prefer to 
 focus on the $n_e$ determination which is independent of this and which implies a higher
 ionization parameter for this galaxy relative to nearby objects.

More immediately, the electron density is related to the pressure in the \textsc{H ii} region through $P = n_e k T_e$. The electron temperature is expected to be set by heating of the ionizing source (in particular its spectral shape) and by cooling of heavy metals and because by our definition in selecting the low-$z$ sample (similar SFR and mass) these two should be similar between the 8 o'clock arc and the low-$z$ sample we expect that the electron temperature should be the same for them. We also calculated the electron temperature based on the CL01 fitting for local galaxies ($T_{e4}= Ê0.62^{+0.06}_{-0.07}$) and the 8 o'clock arc ($T_{e4}=0.67^{+0.08}_{-0.05}$) and those values are consistent with each other. Since the electron temperature in the \textsc{H ii} region is very similar in the low-$z$ sample and the 8 o'clock arc, we conclude that the pressure in the  \textsc{H ii} regions in the 8 o'clock arc is approximately five times higher than that in the typical SDSS comparison galaxy. 

The reason for this elevated \textsc{H ii} region pressure is less clear. \citet{dopita06b} showed that for expanding \textsc{H ii} regions the ionization parameter depends on a number of parameters. A particularly strong dependence was seen with metallicity, but as our comparison sample has similar
metallicity to that of the 8 o'clock arc we can ignore this. The two other major effects on the 
ionization parameter come from the age of the \textsc{H ii} region and the pressure of the surrounding ISM. It is of course possible that we are seeing the 8 o'clock arc at a time when its \textsc{H ii} regions have very young ages, and hence high ionization parameter, relative to the local comparison sample. Since we are considering very similar galaxies in terms of star formation activity and 
probe a fairly large scale this seems to be a fairly unlikely possibility, but it cannot be excluded for a single object. The pressure in the surrounding ISM in \cite{dopita06a} models has a fairly modest effect on the ionization parameter, $U\propto P_{\mathrm{ISM}}^{-1/5}$. Thus we would expect the ISM density in the 8 o'clock arc to also be higher than that in the comparison
sample by a factor of $\sim 5$. We already saw that the gas surface density is higher than that in the comparison sample by a factor of $>2$ thus this is not an unreasonable result and it does not seem to be an uncommon result for high-$z$ galaxies \citep[][]{shirazi13}.

\section{Source Reconstruction}
\label{sec:source}

In order to study emission line maps and the kinematics of the galaxy in the source plane, 
we need to reconstruct the morphology of the 8~o'clock arc using gravitational lens modelling. 
The lens modelling also allows us to derive the magnification factors of the 
multiple-lensed images which were  used to estimate the corrected SFR and the stellar mass 
of the galaxy in the previous section. In the following, we describe our lens modelling 
procedure.
 \begin{figure*}
\centerline{\hbox{\includegraphics[ trim=11cm 3cm 7cm 5cm, clip=true,width=0.9\textwidth, angle=0]
             {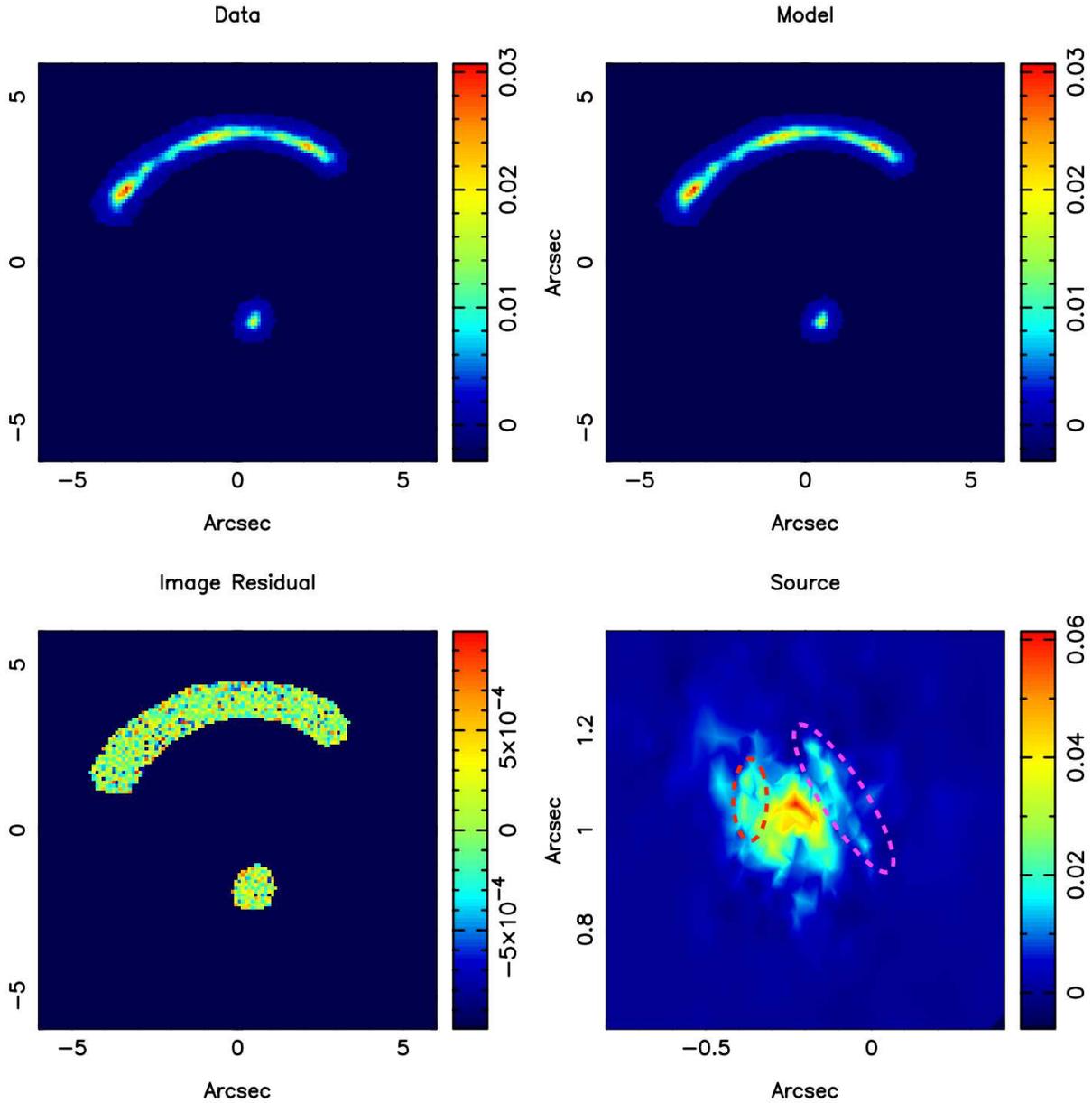}}}
        \caption{Top-left panel: the arc and the counter image in the $B$ band $HST$ image. 
         The foreground galaxy (lens) has been removed from this image. Top-right panel: the best-fit model.
         Lower-left panel: the residuals after subtracting this model from the data.
          The reconstructed $B$ band $HST$ image is shown in the lower-right panel. From this image we 
	see that the source in the rest frame UV consists of at least three components: the main galaxy 
	component, a clump separated by 0.15 arcsec, which is shown by the purple dashed ellipse and 
	another clump separated by 0.15 arcsec, which is shown by the red dashed ellipse.}
 \label{fig:reconstruction-450w}
\end{figure*}
%

\subsection{Gravitational lens modelling}

To reconstruct the lens model for this system, we make use of the Bayesian grid based lens modelling 
technique presented by \citet{vegetti09}, which is optimized for pixelized source surface 
brightness reconstructions.
This technique is based on the optimization of the Bayesian evidence, which is given by a combination of the $\chi^2$ and a source regularization term. The $\chi^2$ is related to the difference between the data and the model, while the regularization term is a quadratic prior on the level of smoothness of the source surface brightness distribution and is used to avoid noise fitting.

In order to obtain a robust lens model, we first consider the high resolution and high SN ratio 
$B$ band $HST$ image. We assume the lens mass distribution to follow a power-law elliptical profile with 
surface mass density, in units of the critical density, defined as follows
\begin{equation}
k(x,y) = \frac{k_0 \;(2-\frac{\gamma}{2}) \; q^{\gamma-3/2}}{2(q^2 \;((x-x_0)^2 +{r_c}^2)+(y-y_0)^2)^{(\gamma-1)/2}}
\label{eq2}
\end{equation}

We also include a contribution from external shear. In particular, the free parameters of the model are the mass 
density normalization $k_0$, the position angle $\theta$, the mass density slope $\gamma$ ($\gamma=2.0$ for an 
isothermal mass distribution), the axis ratio $q$, the center coordinates $x_0$ and $y_0$, the external shear strength 
$\Gamma$, the external shear position angle $\Gamma_{\theta}$, and the source regularization level (i.e., a measure 
of the level of smoothness of the source surface brightness distribution). The core radius is kept fixed to the 
negligible value of $r_c \equiv 0$, since Einstein rings only  make it possible to constrain the  mass distribution at the Einstein radius. 

The most probable parameters of the model, i.e., the parameters that maximize the Bayesian evidence, are $k_0 = 3.367$,  $\theta = 14.54$, $q =0.618$, $\gamma = 2.009$, 
$\Gamma$ = 0.062 and $\Gamma_{\theta} = 10.597$. Using the same analytic mass model of Equation \ref{eq2} for the lens galaxy, we also model the 
NICMOS data. While the $B$ band $HST$ data probe the rest frame UV and have a higher resolution in comparison to 
the NICMOS data, the latter have the advantage of providing us with information about the continuum in the $J$ and $H$ 
bands, where \hb\ and $\oii$ emission lines are located in the spectra.
The most probable mass model parameters for the NICMOS data are $k_0 =3.328$, 
$\theta =14.30$, $q=0.672$, $\gamma= 2.020$, $\Gamma =0.077$ and $\Gamma_{\theta}=13.37$. Both results are consistent
 with DZ11 best-fit parameters, within the error bars and are consistent with each other. The difference between the mass model parameters gives a good quantification of systematic errors ($\sigma_{K_0}=0.039$, $\sigma_{\theta}=0.24$, $\sigma_{q}=0.054$, $\sigma_{\gamma}=0.018$, $\sigma_{\Gamma}=0.015$, $\sigma_{\Gamma_{\theta}}=2.773$).
These most probable parameters are used to map the image plane into the source plane and reconstruct the original 
morphology of the 8~o'clock arc in the $B$ and $H$ bands, respectively. 

Thanks to the Bayesian modelling technique, the most 
probable source surface brightness distribution for a given set of lens parameters is automatically provided. 
The reconstructed $B$ band $HST$ image is shown in the lower-right panel of Fig. \ref{fig:reconstruction-450w}. From this image, we 
can see that the source in the rest frame UV consists of multiple components, including the main galaxy component 
and two clumps separated by 0.15\,arcsec (i.e., 1.2\,kpc in projected distance) indicated by the purple and red ellipses. 
The reconstructed $H$ band $HST$ image is shown in Fig. \ref{fig:reconstruction-zoom160w}. This image shows
 that the source in the rest-frame optical also consists of multiple components. 
 
 From the modelling the HST data, we get a set of best lens parameters and source regularization level that maximize the Bayesian evidence. That means we used the same analytic model and re-optimized for both the source regularization and lens parameters ($k_0$, $\theta$, etc.). For modelling the SINFONI data, however because the SN is poor we keep the lens parameters fixed, but we re-optimise for the source regularization level.  

%
	\begin{figure}
	\centerline{\hbox{\includegraphics[width=0.45\textwidth, angle=90]
             {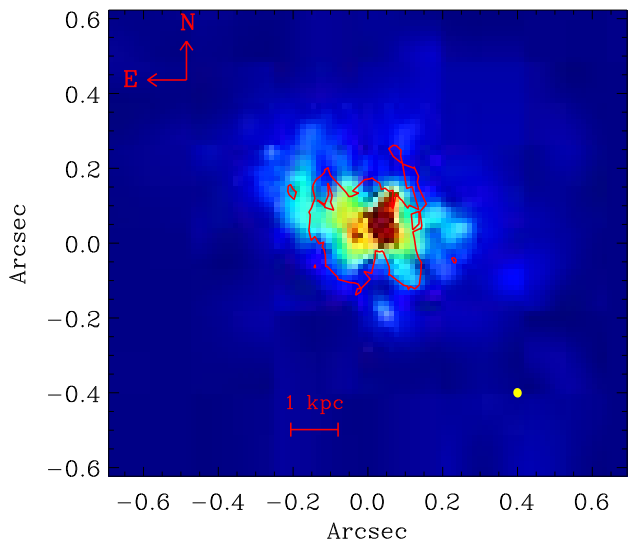}}}
        \caption{The reconstructed $H$ band $HST$ image is shown in the plot. From this image we 
	can see that the source in the rest frame optical is formed of multiple components and two main galaxy components. 
	The  clump in the reconstructed UV image is partially resolved in this image and it shows a slightly different morphology. 
	The contour shows the $HST$ $B$-band reconstructed image. The yellow ellipse in the corner of the image shows the spatial 
	resolution in the source plane.}
 	\label{fig:reconstruction-zoom160w}
	\end{figure}

	\begin{figure}
	\centerline{\hbox{\includegraphics[width=0.47\textwidth, angle=90]
          {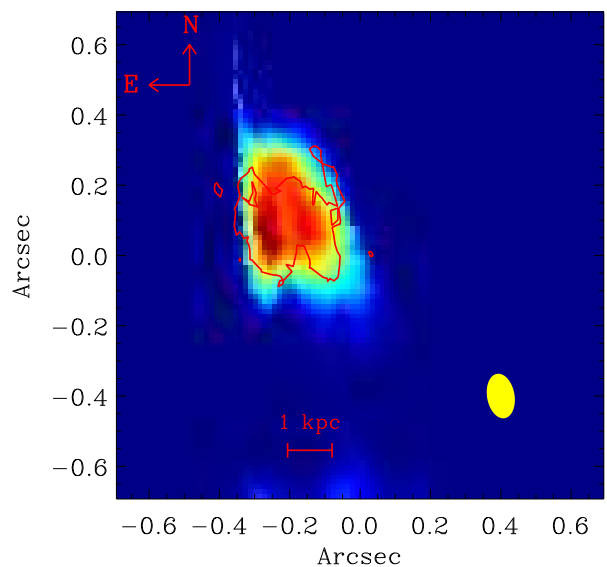}}}
      	  \caption{The reconstructed \hb\ image is shown in the plot. From this image we 
	can see that the source in the rest frame optical is formed of multiple components and two main galaxy components. 
	The  clump in the reconstructed UV image is not resolved in this image. 
	The contour shows the reconstructed $HST$ $B$-band image from Fig. \ref{fig:reconstruction-450w}.
	 The yellow ellipse in the corner of the image shows the spatial resolution in the source plane.}
 	\label{fig:recons-hb}
	\end{figure}
	\begin{figure*}
	\centerline{\hbox{\includegraphics[ trim=2cm 0cm 8cm 0cm, clip=true,width=0.37\textwidth, angle=90] {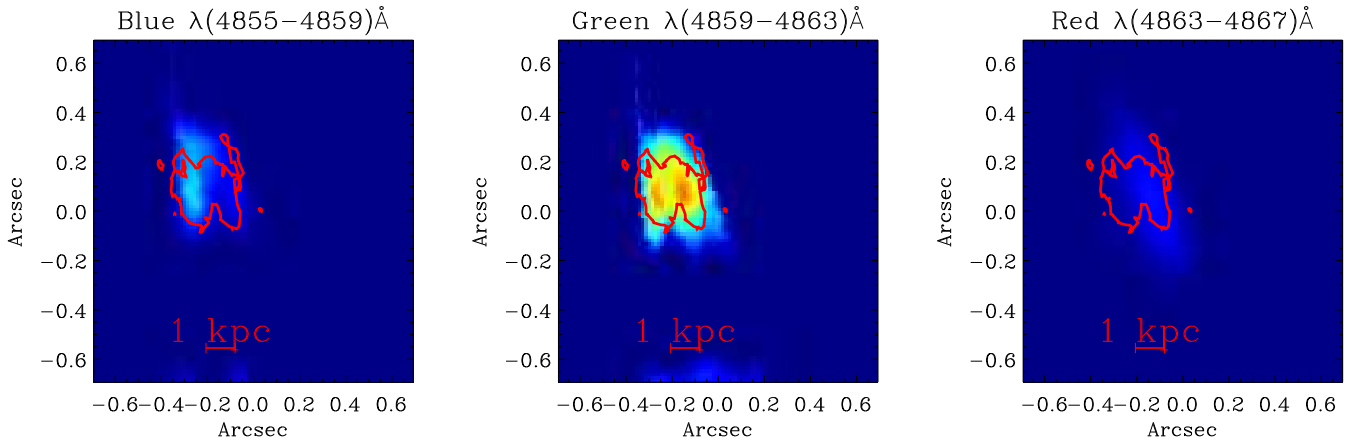}}}
         \caption{From left to right this figure shows reconstructed \hb\ maps from the blue, green and red components of the spectral line.
         The contour line shows the reconstructed $HST$ $B$-band image from Fig. \ref{fig:reconstruction-450w}. It can be seen that the blue map predominantly contributes 
         to the eastern part --- from a detailed inspection of the lens model we find that in the 
         image plane this is predominantly seen in the A1 and A2 images.  Note also that the 
         red component of the spectral line predominantly originates in the west. }
 	\label{fig:bgr-rec}
	\end{figure*}
\subsubsection{Reconstructed-\hb\ and $\oii$ emission line maps in the source plane}

Here we make use of the $B$ band $HST$ data modelling to reconstruct the \hb\ and $\oii$ emission 
line maps of the galaxy. These lines have the highest SN that we obtain
from the SINFONI data. 

For each spectral pixel image (frame) of the SINFONI data cubes, we derive the most probable  
source surface brightness distribution by keeping the lens parameters fixed at the best values 
recovered from the $B$ band $HST$ data modelling (after taking into account the rotation of the image), 
while re-optimizing for the source regularization level. Because of the relatively low SN 
SINFONI data and non-homogeneous sky background, we can not use all 
the lensed images. We focus instead on the highest magnification image, which is the A2 image. 

Before reconstructing the \hb\ map in the source plane, we first bin in the spectral direction by a 
factor of 4. This corresponds to the spectral resolution (FWHM= 7.9 \AA) that we measure from the 
line widths of the night sky lines around the \hb\ line. This provides us with higher SN 
image plane frames. We finally make a \hb\ source cube from these reconstructed source frames 
and use that to derive the kinematics of the galaxy.
A reconstructed \hb\ map is shown in Fig. \ref{fig:recons-hb}. This image also shows two galaxy components. 
In order to better understand the morphology of the \hb\ image, we divide the \hb\ spectral range into 
three equal spectral bins defined as blue ($\lambda(4855-4859)$\AA), green ($\lambda(4859-4863)$\AA) 
and red ($\lambda(4863-4867)$\AA) intervals, corresponding to three SFR maps shown in Fig. \ref{fig:bgr}. 
We then reconstruct the source surface brightness distribution for each of these images, using the same 
method as was used for the whole \hb\ image. The three panels in Fig. \ref{fig:bgr-rec} show the reconstructed 
sources for these images. We can see that the west part of the \hb\ line map is very weak and only 
dominates in the red image (right-hand panel); on the other hand, the eastern part is dominant in the blue image (left-hand panel). 
Fig. \ref{fig:reconstruction-oii} shows a reconstructed $\oii$ image of the galaxy. We see that the eastern component is 
dominant in this image. Here we are unable to separate the two components; this might be due to a higher 
$\oii/\hb$ ratio in the eastern component, but could also be caused by the lower spatial resolution at these wavelengths. 
However, we rule out the latter explanation by convolving the \hb\ map with a Gaussian PSF matching the slightly different \oii map ($J$-band) PSF. 

\begin{figure}
\centerline{\hbox{\includegraphics[width=0.5\textwidth, angle=90]
            {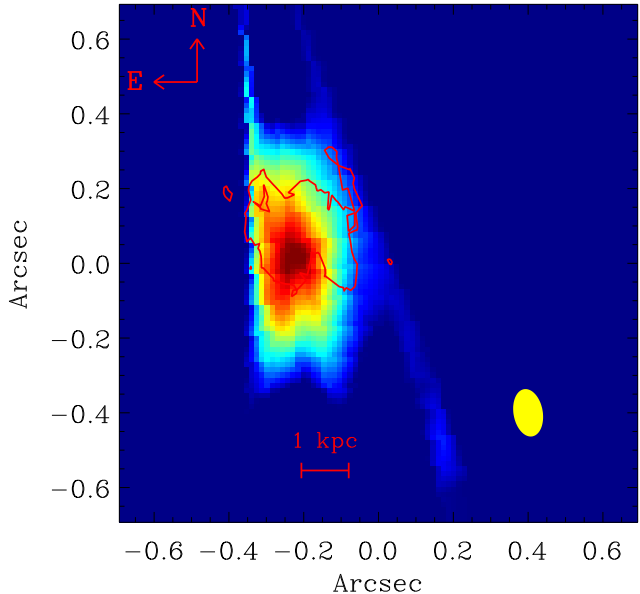}}}
      \caption{The reconstructed $\oii$ image. Here we are unable to separate two components; 
      this might be due to a higher $\oii/\hb$ ratio in the left component. The red contour shows the reconstructed $HST$ $B$-band image from Fig. \ref{fig:reconstruction-450w}.} 
 \label{fig:reconstruction-oii}
\end{figure}
%

\subsection{\hb\ profile of the reconstructed source}
\label{sec:profile-reconstructed}

We use the same fitting method as we used in Section \ref{sec:analysis} to fit a two component 
Gaussian to the \hb\ profile for every pixel. We also integrate over the total flux of the galaxy 
and fit a two component Gaussian to it to compare it to our study in the observed plane (see Section \ref{sec:profile}).
We carry out these fits using the {\scriptsize{MPFIT}} package in {\scriptsize{IDL}}\footnote{http://cow.physics.wisc.edu/$\sim$craigm/idl/mpfittut.html}. During the fitting, we require the lines to have the same velocity widths. Fig. \ref{fig:source-hb-profile} shows the \hb\ profile derived from the reconstructed \hb\ source. 
We see that this profile also retains two components. These Gaussian components are shown by red dashed lines in Fig. \ref{fig:source-hb-profile}. The width of the Gaussian for both components 
is 1.59 \AA, which gives a velocity dispersion of $98\pm44$ $\rm{km\; s^{-1}}$. 
This is consistent, within the errors, with our estimated velocity dispersion for the A2 and A3 images 
in the observed plane (i.e., $104\pm42\;\rm{km\; s^{-1}}$). 
The velocity offset between the two components is $246\pm46\; \rm{km\; s^{-1}}$ and matches the offset that 
we derive for the A2 and A3 images in the image plane. 
 Fig. \ref{fig:source-velocity} shows the velocity and velocity dispersion maps derived from the 
 reconstructed \hb\ source. From these maps, we see that the eastern galaxy component has a lower velocity 
 and velocity dispersion relative to the western component. The western component also shows a smoother 
 velocity gradient.
 
 The \hb\ line flux divided by the $H$-band continuum flux is shown as a proxy for EW(\hb) in 
 Fig. \ref{fig:eqw}. Here, the $H$-band continuum map is convolved to the same PSF as the \hb\ map.
 From this we see that the outskirts of the galaxy show a clumpy and higher EW(\hb). The eastern component of the galaxy 
 also shows a higher EW(\hb), which might be interpreted as a younger age relative to that of the main component.
 
\begin{figure}
\centerline{\hbox{\includegraphics[trim=3cm 13cm 1cm 0.5cm, clip=true,width=0.55\textwidth, angle=90]
             {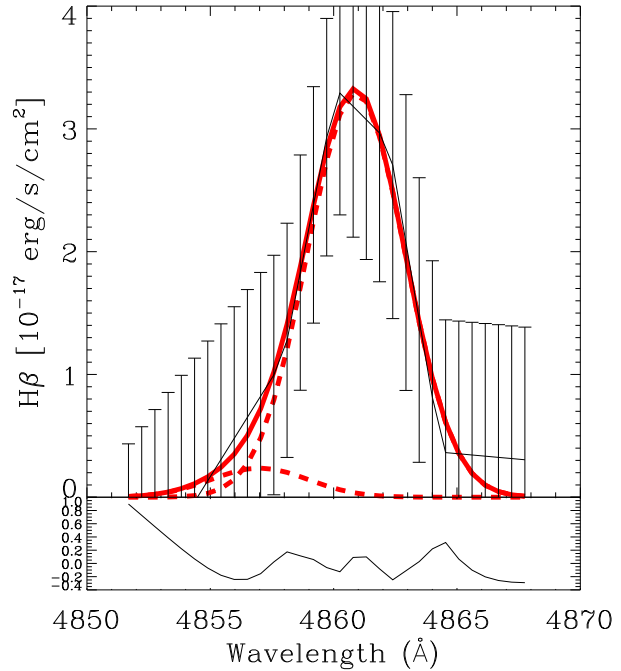}}}
         \caption{\hb\ profile derived from the reconstructed \hb\ source. 
         The width of both Gaussian components is 1.59 \AA, which gives a velocity dispersion 
         of 98 $\rm{km\; s^{-1}}$. The velocity offset between the two components is 246 $\rm{km\; s^{-1}}$. In 
         the source plane, we have fewer bins than Fig. \ref{fig:spec-A2-A3}, since we bin in 
         spectral resolution by a factor of 4 before reconstructing the \hb\ map in the source plane. 
          To make this profile, we interpolate between those bins to have the same binning as in Fig. \ref{fig:spec-A2-A3}. The lower panel shows the residuals if we fit only one Gaussian component to the profile.}
 \label{fig:source-hb-profile}
\end{figure}

\begin{figure}
\centerline{\hbox{\includegraphics[width=0.4\textwidth, angle=90]
             {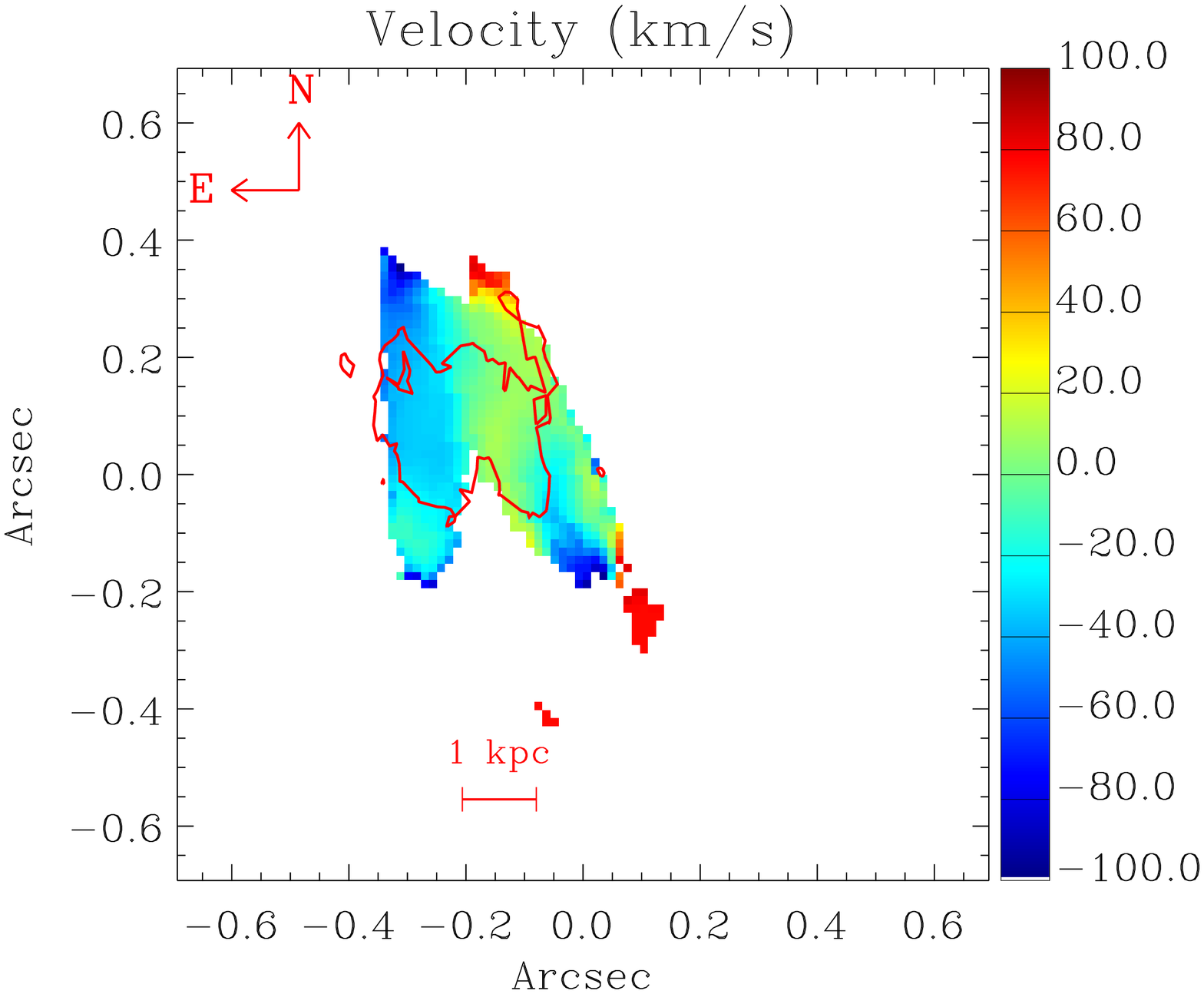}}}
             \centerline{\hbox{\includegraphics[width=0.4\textwidth, angle=90]
             {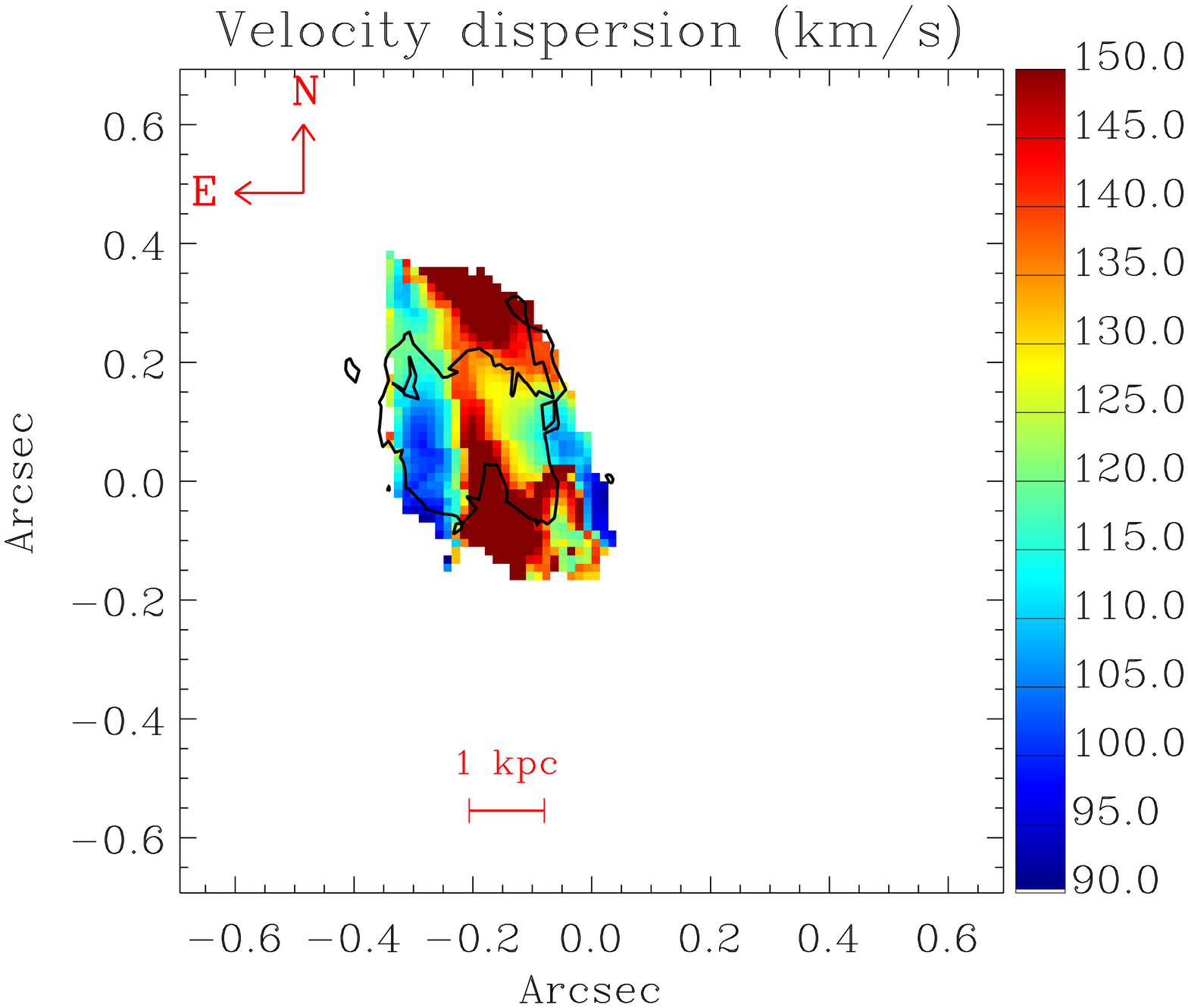}}}
              \centerline{\hbox{\includegraphics[width=0.4\textwidth, angle=90]
             {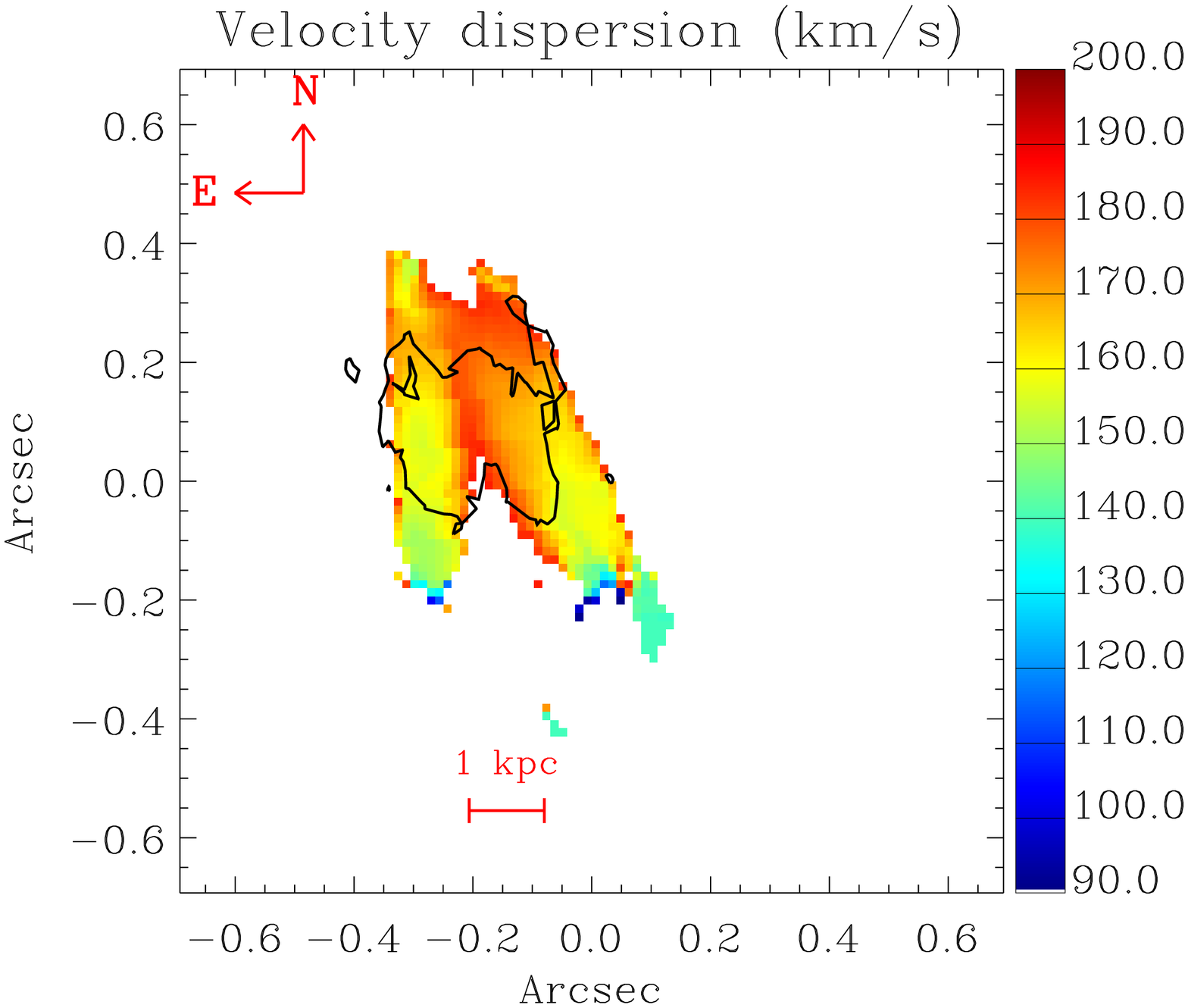}}}
         \caption{The top and middle panels show the velocity and velocity dispersion 
         maps, respectively. The velocity map is derived using single Gaussian fits but velocity dispersion map represents both components. These are derived from the reconstructed \hb\ source. The bottom panel shows the velocity dispersion map derived using single Gaussian fits. The contours show the reconstructed $HST$ $B$-band image from Fig. \ref{fig:reconstruction-450w}.}
 \label{fig:source-velocity}
\end{figure}

\begin{figure}
\centerline{\hbox{\includegraphics[ width=0.35\textwidth, angle=90]
             {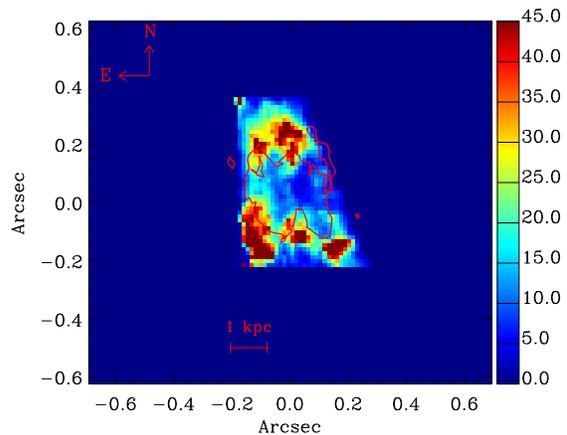}}}
                     \caption{\hb\ line flux divided by the $H$-band continuum flux shown as a proxy for EW(\hb). We see that 
                     the outskirts of the galaxy show a clumpy and higher EW(\hb). The eastern component of the galaxy 
                     shows a higher EW(\hb), i.e., a younger age, relative to that of the main component. The red contour shows the reconstructed $HST$ $B$-band image.}
 \label{fig:eqw}
\end{figure}

\section{Dynamics}
\label{sec:dynamic}
\subsection{\hb\ kinematics}
\label{sec:kinematics}
To test whether the kinematics of the galaxy are consistent with those of a rotating disk, we compare the 
velocity and velocity dispersion maps derived from the reconstructed \hb\ map with an exponential 
disk model. Given the low resolution and the low SN of our data, we simulate a very
simple system.
The disk models is created using the {\scriptsize{DYSMAL}}  {\scriptsize{IDL}} code \citep[][see also Cresci et al. 2009 
for description of the code]{davies11}. 
The code was used extensively to derive intrinsic properties of disk galaxies 
\citep[e.g., for estimating the dynamical mass of high-$z$ galaxies; see ][]{cresci09}. 
The code uses a set of input parameters which constrain the radial mass profile as well as 
the position angle and systemic velocity offset, in order to derive a 3D data cube with one 
spectral (i.e., velocity) and two spatial axes. This can be further used to extract kinematics. 
The best-fit disk parameters are derived using an optimized $\chi^2$ minimization routine 
and the observed velocity and velocity dispersion. 
The mass extracted from {\scriptsize{DYSMAL}} is that of a thin disk model assuming 
supported only by orbits in ordered circular rotation. 

We do not have any constraints on the inclination of our system. Therefore, we use a nominal 
inclination of 20 degrees. We account for spatial beam smearing from the PSF
and velocity broadening due to the finite spectral resolution of the instrument and also rebin by a factor of 4 in the spectral direction
in our modelling. We then compare this spatially
and spectrally convolved disk model to the observations.

We focus on the western component in the velocity map shown in Fig. \ref{fig:source-velocity} because 
from the lens modelling we know that this part contains the main component of the galaxy and also shows 
a smoother observed gradient.
The best-fitting exponential disk model for this component is shown in Fig. \ref{fig:source-velocity-model}. We show the observed velocity derived from the reconstructed \hb\ map along the slit (shown by dashed lines in Fig. \ref{fig:source-velocity-model}) in Fig. \ref{fig:source-velocity-model-trend}.
While the disk model can reproduce some large-scale features of the velocity field, the residuals 
are substantial. We can therefore rule out a single rotating disk as a reasonable description of 
this system. We conclude that the 8 o'clock arc has a complex velocity field that cannot be explained by a simple rotating disk.

Furthermore, there appears to be a second component from a clump (see the red ellipse in Fig. \ref{fig:reconstruction-450w}) 
that partially overlaps with this component. 
Whether this is a sign of an on-going merger is difficult to ascertain with the present data. Indeed, the SN in 
\hb\ does not warrant a much more complex model to be fitted.
%
\begin{figure}
\centerline{\hbox{\includegraphics[ trim=2cm 3cm 3cm 2cm, clip=true,width=0.37\textwidth, angle=90]
             {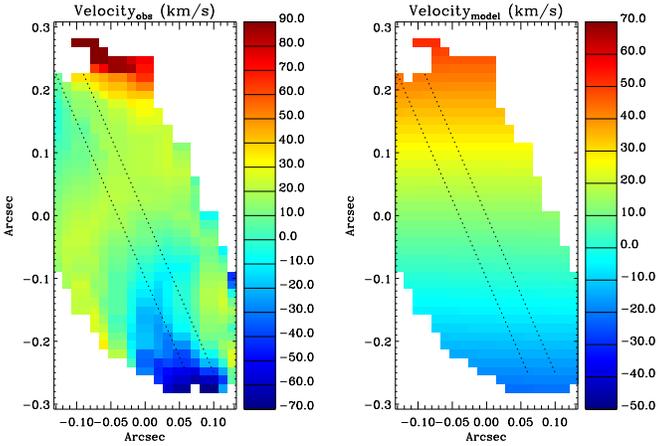}}}
                     \caption{The left-hand panel shows the observed velocity derived from the reconstructed \hb\ map. 
                     The right-hand panel shows the best velocity fit.}
 \label{fig:source-velocity-model}
\end{figure}

\begin{figure} 
\centerline{\hbox{\includegraphics[ width=0.45\textwidth, angle=180]
             {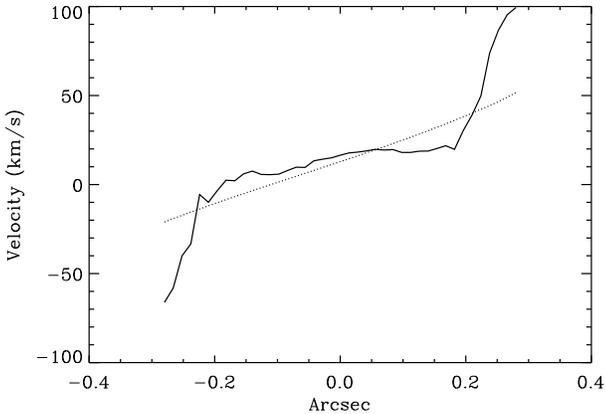}}}
                     \caption{The observed velocity derived from the reconstructed \hb\ map along 
                     the slit (shown by dashed lines in Fig. \ref{fig:source-velocity-model})
                     is shown by the solid curve. The dashed line here shows the 
           		 best-fit velocity model.}
 \label{fig:source-velocity-model-trend}
\end{figure}

\subsection{Dynamical mass}
\label{sect:dynmass}

DZ11 estimated the dynamical mass of the 8 o'clock arc from the line widths via the relation presented by \citet{erb06b}.
We use the same method to estimate the dynamical mass using our estimated velocity dispersion ($\sigma$) and half-light radius.
For rotation-dominated disks, DZ11 assumed that the enclosed dynamical 
mass within the half-light radius, $r_{1/2}$, is $M_{\rm dyn,rot}\,(r<r_{1/2}) 
= (2.25 \sigma^2 r_{1/2})/G$ and multiply this resulting mass by 2 to 
obtain the total dynamical mass, where $G = 4.3\times 10^{-6}\,\rm kpc\,(km\,s^{-1})^2\;M_{\sun}^{-1}$ is the 
gravitational constant.
For dispersion-dominated objects, they applied the 
isotropic virial estimator with $M_{\rm dyn,disp} = (6.7 \sigma^2 r_{1/2})/G$, 
appropriate for a variety of galactic mass distributions \citep{binney08}. In 
this case, $M_{\rm dyn,disp}$ represents the total dynamical mass.

For estimating the half-light radius we run \textit{galfit} on the reconstructed $B$ and $H$ band images. This gives us 
$r_{1/2}=2.8\pm0.2$ kpc.
We measure $\sigma=104\pm42$ $\rm{km\; s^{-1}}$ and a rotation-dominated dynamical mass ${\log \; (M_{dyn}/M_{\sun}})
= 10.2\pm0.3$ and a dispersion-dominated dynamical mass 
${\log\; (M_{dyn}/M_{\sun}})= 10.7\pm0.27$ (these values are corrected for instrumental broadening).
Using the de-lensed spectra, we also estimate $\sigma=98\pm44$ $\rm{km\; s^{-1}}$, which give us 
$10.1\pm0.6$, and  $10.6\pm0.6$ for the rotation-dominated and dispersion-dominated dynamical masses, respectively. 
The disk model fit can also provide a dynamical mass estimate, ${\log\;(M_{dyn}/M_{\sun}})= 9.5$, but 
we do not use this here because it only accounts for the west component of the velocity map. 

We note that the idea of using a single line width to estimate dynamical mass is not convincing 
(this should be done for blue, green and red components individually). In that case the velocity 
field is clearly more like a merger, so neither of these dynamical mass indicators are reliable. 
Therefore, obtaining a robust dynamical mass estimate would require considerably more sophisticated models. 
The simple models are not physically constraining.


\subsection{A massive outflow of gas?}
\label{sec:outflow}
It has been shown that many of high-$z$ star-forming galaxies show evidence for powerful galactic outflows, indicated by studying UV absorption spectroscopy
 \citep[][]{pettini00,shapley03, weiner09, steidel10, kornei12} and broad \ha\ emission-line profiles
\citep{shapiro09,genzel11, newman12a}. Recently \citet{newman12b}
showed how galaxy parameters (e.g., mass, size, SFR) determine the strength of these outflows. They decomposed the emission line profiles into broad and narrow components and found that the broad emission is spatially extended over $\sim$ a few kpc. \citet{newman12b} showed that the star formation surface density enforces a threshold for strong outflows occurring at $1\; \rm{M_{\odot} \; yr^{-1} \; kpc^{-2}}$. The threshold necessary for driving an outflow in local starbursts is $0.1\; \rm{M_{\odot} \; yr^{-1} \; kpc^{-2}}$ \citep{heckman02}. 

The 8 o'clock arc with integrated star formation surface density of $9.2 \; \rm{M_{\odot} \; yr^{-1} \; kpc^{-2}}$ is certainly in the regime of strong outflow. If we consider the ratio of the Gaussian flux in the blue-shifted component to that of the main component ($\sim0.5$) as the $F_{broad}/F_{narrow}$, then our result is consistent with what \citet{newman12b} show in their Fig. 2. However, we note that this definition is not exactly what \citet{newman12b} introduced as $F_{broad}/F_{narrow}$ as we do not fit broad and narrow components but two component with the same width. 

In our data we find a blue-shifted component to \hb\ as discussed in Section~\ref{sec:profile}. As we mentioned earlier we used the same Gaussian width for both main and blue-shifted components. Given the low SN, a unique broad fit with a physical meaning can not be found considering the fact that residual from sky lines might create broad line widths. The velocity offset between this component and the main
component of the \hb\ line is $\approx 190$ $\rm{km\; s^{-1}}$ for the A2 image and $\approx 280$ $\rm{km\; s^{-1}}$ for the A3 image. This blue-shifted component could be due to an outflow of gas or a minor merger. 

In support of the outflow picture, \cite{finkelstein09} and DZ10 both observed that ISM lines in 
the rest-UV spectrum of the 8 o'clock arc are blue-shifted relative to
the stellar photospheric lines. They also argued that this was a sign of an outflow of gas from the galaxy and taken together with the
SINFONI results this strengthens the outflow picture. A further argument for an outflow is the fact
that as we saw in Section~\ref{sec:density}, the \textsc{H ii} regions have an elevated 
internal pressure, at least compared to similar galaxies locally, and it is reasonable to assume 
that this aids in driving an outflow \citep{heckman90}. 

However, based on DZ10 results, it is noticeable that the strong absorption lines reach $\sim$ zero intensity at $v=-200 \;\rm{km \;s^{-1}}$ indicating that the outflowing gas entirely covers the UV continuum, whereas the blueshifted \hb\ is spatially separated from the majority of the rest-UV emission (e.g. in Fig. 16).

We note that the analysis of UV lines in DZ10 is done in the image plane and only applies to the region sampled by the long-slit. Within this region the UV lines do appear to show that the outflowing gas entirely covers the UV continuum. However it is not possible to compare this directly to our results because of this limitation. Indeed it is clear from our analysis in the source plane that the blue-shifted component \hb\ comes from a spatially separated region. These two results can be easily reconciled if the slit spectrum of DZ10 predominantly samples in the image plane the region where the blue-shifted \hb\ originates. Spatially variable dust attenuation could complicate this scenario and without spatially resolved spectroscopic UV data we cannot make a reliable comparison.

We expect also broader line width for an outflowing component \citep[e.g.,][and local ULIRG outflows]{newman12b}. The reason why we do not see a broader blue-shifted component is because of our method for line fitting and was justified above given our low SN data.

Based on a single Gaussian fit to the \hb\ profile, the line width (FWHM) that we derive for the 8 o'clock arc is $330\pm80\; \rm{km \;s^{-1}}$. The properties of the 8 o'clock arc (SFR, integrated star formation surface density and stellar mass ) are in the range of high bins defined by \citet{newman12b} and the line widths presented by \citet{newman12b} for galaxies showing significant outflows are $510\pm12$, $500\pm 16$ and $520\pm11\;\rm{km \;s^{-1}}$, respectively. These are broader than what we find but we expect that a significant part of this is due to us fitting a single Gaussian to the \hb\ line as compared to what can be done with \ha.  

It could be useful to know whether emission line ratios of the blueshifted component are consistent with shocked gas (as expected for an outflow) as opposed to photoionization in \hii\ regions. It was suggested by \citet{letiran11} that it is not expected to detect the line emission from shocked gas as its emission line surface brightness is too faint and it is expected to be hidden by the emission caused by photoionising radiation from the starburst. However, \citet{yuan12} and \citet{newman13} showed that the detection of shock induced emission might be possible, especially the integral field spectra can help considerably as the blueshifted component is separated spatially and spectrally from the bulk of star formation. However, we can not distinguish shocked versus photonionzed emission based on the current data. This could be done with sufficiently deep IFU data of the required line ratios.

In support of the merger picture, Fig. \ref{fig:bgr-rec} indicates that much of the
blue component flux is extended and arguably galaxy shaped. Obtaining sufficiently deep IFU spectra of the separated blueshifted component can help considerably to study the origin of this component.

Taking the evidence from the UV ISM lines, the \hb\ profile, the lens model and the high 
ISM pressure together, it is likely that there might be a significant outflow component to the blue-shifted wing,
but the question here is whether these observations correspond to a reasonable amount of gas in the outflow. 
To answer this we also estimate the mass of ionized hydrogen from the
luminosity of \ensuremath{\mathrm{H}\beta},
$L_{\ensuremath{\mathrm{H}\beta}}$, of the blue-shifted component and main component of the 8 o'clock arc using
\citep[e.g.,][]{dopita03}:
\begin{equation}
  \label{eq:m_ionised}
  M_{\mathrm{ionized}} = \frac{m_\mathrm{H} \times L_{\ensuremath{\mathrm{H}\beta}}}{1.235\times
    10^{-25} T_4^{-0.86} n_e},
\end{equation}
where $m_\mathrm{H}$ is the mass of the hydrogen atom, $T_4$ the
electron temperature in units of 10,000\,K and $n_e$ the electron
density. We use $T_4=0.67$ that we derived from the CL01 fitting.
The mass of ionized hydrogen for the blue-shifted component with 
$L_{\ensuremath{\mathrm{H}\beta}}$ of $3.5\times 10^{41} \mathrm{erg} \; \mathrm{s}^{-1}$ is 
$10^{6.45\pm0.1}$ $M_{\sun}$ assuming $n_e=600 \;\rm{cm^{-3}}$, which can be contrasted with the mass of ionized hydrogen in the 
main component which is $10^{6.91\pm 0.1} M_{\sun}$ (with $L_{\ensuremath{\mathrm{H}\beta}}$ = $9.8
\times 10^{41} \mathrm{erg} \; \mathrm{s}^{-1}$). The estimate for the blue-shifted 
component is clearly an upper limit because from the argument above it seems likely that the 
blue component is not purely an outflow. Thus, taking this at face value, we would find that with an
outflow rate of 10\% of the SFR, a conservative value given local observations
\citep[e.g.,][see also Martin12]{Martin99,martin05,Martin06}, we need the current star formation activity to have lasted $< 3\times 10^8 \; yr$, which is not unreasonable
(to estimate the star formation time-scale, the mass is divided by 10\% of the SFR).

In order to use the outflow mass quantitatively, we need to have an estimate of outflow mass in neutral hydrogen. However, we note that we 
can not estimate this for the blue-shifted component since $N_{\rm{HI}}$ estimated in DZ10 is given for the whole galaxy.
To estimate the total neutral hydrogen mass in the outflow and the mass
outflow rate \citep{pettini00}, we use the following formulae given in \cite{Verhamme08} 
\begin{equation}
  \label{eq:m_neutral}
  M_{\mathrm{HI}} \approx 10^7 \; \Bigl(\frac{r}{1\; \rm{kpc} }\Bigr)^2 \Bigl(\frac{N_{\rm{HI}}}{10^{20} \rm{cm}^{-2}}\Bigr) \; M_{\odot},
\end{equation}
\begin{equation}
  \label{eq:outflow_rate_neutral}
  \dot{M}_{\mathrm{HI}} = 6. \; \Bigl(\frac{r}{1\; \rm{kpc} }\Bigr) \Bigl(\frac{N_{\rm{HI}}}{10^{20} \rm{cm}^{-2}}\Bigr) \Bigl(\frac{V_{exp}}{200 \; \rm{km\; s^{-1}}}\Bigr) \; M_{\odot} \; yr^{-1},
\end{equation}

where the first equation relate \hi\ mass in the shell to its column density $N_{\rm{HI}}$ and the second equation assumes that the mechanical
energy deposited by the starburst has produced a shell of swept-up interstellar matter that is expanding with a velocity of $V_{exp}$.

Assuming our estimated half-light radius from the rest-frame UV reconstructed image and our assumed outflow velocity ($r = 2.8$ kpc, $V_{exp}=200\; \rm{km\; s^{-1}}$) and taking $N_{\rm{HI}}=10^{20.57}\; \rm{cm}^{-2}$ from DZ10, we find neutral gas masses of $M_{\mathrm{HI}}=2.9\times10^8 M_{\odot}$ and an outflow rate of $\dot{M}_{\mathrm{HI}}=62.4 \;M_{\odot} \; yr^{-1}$. This gives us a mass-loading factor of $\eta=\dot{M}_{\mathrm{HI}}/\rm{SFR}=0.27$. This inferred mass-loading factor of \hi\ is small compared to those measured by \citet{newman12b} for galaxies with similar SFR surface densities to the 8 o'clock arc. We note that we measure a smaller $V_{exp}$ and if we consider two times higher $V_{exp}$ then we infer a higher mass-loading factor. \citet{newman12b} results are derived for the broad flux fraction while we infer the mass-loading not for a broad component. If we consider the ratio of the Gaussian flux in the blue-shifted component to that of the main component ($\sim0.5$) as the $F_{broad}/F_{narrow}$ then we infer a mass-loading factor of $\sim 1$ which is consistent with their results considering the uncertainty in the mass-loading factor.

\section{Conclusions}
\label{sec:conclusion}
We present a spatially-resolved analysis of the 8 o'clock arc using NIR IFU data. From this we recover 
the \hb\ map and the spatially-resolved \hb\ profile. 
We showed that \hb\ has different profiles at different spatial pixels and is composed of multiple components. We carefully modelled the strong emission lines in the galaxy and compared the 
results to a local comparison sample. This allowed us to conclude the following.
\begin{itemize}
\item The 8 o'clock arc lies on the same $M_*$-O/H-SFR manifold as similar star-forming galaxies locally \citep[][DZ11]{lara-lopez10,mannucci10}.
\item The gas surface density in the 8 o'clock arc $\log (\Sigma_{\mathrm{gas}}/\;\rm{M_{\odot}\;pc^{-2}})= 1.6$ (1-$\sigma$ range of 1.46-1.87) is more than twice ($\times 2.16$ with 1-$\sigma$ range of 1.55-4.01) that of similar galaxies locally $\log (\Sigma_{\mathrm{gas_{(analogs)}}}/\;\rm{M_{\odot}\;pc^{-2}})= 1.27$ (1-$\sigma$ range of 1.02-1.48). 
Comparing this with other high-$z$ results \citep[e.g.,][who measure gas surface densities in the range of 2.5-3.3 $\rm{M_{\odot}\;pc^{-2}}$]{mannucci09}, the gas surface density for the 8 o'clock arc is lower. Note that as mentioned by \cite{mannucci09}, they are sampling the central, most active parts of the galaxies, so those values should be considered as the maximum gas surface densities.
 
\item The electron density, and thus the \textsc{H ii} region pressure, in the 8 o'clock arc is approximately five times that of similar galaxies locally. As \citet{wuyts12} pointed out, the electron density measurements for high-$z$ galaxies range from the low density limit to $n_e>10^4 \;\rm{cm^{-3}}$. Although these differences depend on the method for measuring the electron density, these also imply a huge difference in the physical properties of star-forming regions in star-forming galaxies at $z\sim2$. The difference between electron density at low-$z$ and high-$z$ has been studied recently by \citet[][]{shirazi13} who compared a sample of 14 high-$z$ galaxies with their low-$z$ counterparts in the SDSS and showed that high-$z$ star-forming galaxies that have the same mass and sSFR as low-$z$ galaxies have a median of eight times higher electron densities.
\end{itemize}
Taken together these results imply that although the 8 o'clock arc seems superficially similar to local galaxies with similar mass and star formation activity, the properties of the ISM in the galaxy are nonetheless noticeably different.

We showed that the two images A2 and A3 have the same \hb\ profiles, which of course is to be expected because they are two images of the same galaxy. But this contrasts with the results from long-slit observations of the object by DZ11 who found different profiles. The similarity of the profiles from the IFU data has allowed us to rule out a significant contribution of substructures to the surface brightness of the A2 image. 

The integrated \hb\ profile of both images show a main component with a blue wing which can be fitted by another Gaussian profile with the same width. The width of the Gaussian components for both images is $1.7\pm0.7$ \AA, which gives a velocity dispersion of $\sim$ $104\pm42\; \rm{km\; s^{-1}}$. 
The velocity offset between the two components is $278\pm 63.5\; \rm{km\; s^{-1}}$ for the A3 image and $191\pm63\;\rm{km\; s^{-1}}$ for the A2 image which are consistent within the errors. 
Since both DZ11 and \cite{finkelstein09} showed that ISM lines are blue-shifted relative to the stellar photospheric lines, 
suggesting gas outflows with 120-160 $\rm{km\; s^{-1}}$, and find a comparatively high pressure in the \textsc{H ii} regions of the 8 o'clock arc, we interpret this blue-shifted component as an outflow. However, we cannot rule out that the blue-shifted 
component might represent a minor merger. 

To study the de-lensed morphology of the galaxy, we used existing $B$ and $H$ band $HST$ images. 
Based on this, we constructed a rigorous lens model for the system using the Bayesian grid based lens modelling 
technique. In order to obtain a robust lens model, we used the lens modelling of the $B$ band $HST$ image to reconstruct the \hb\ line map of the galaxy.
We then presented the de-lensed \hb\ line map, velocity and velocity dispersion maps of the galaxy. 
As an example application we derived the \hb\ profile of the reconstructed source and showed that this also requires two Gaussian components with a width of $98\pm44\;\rm{km\; s^{-1}}$ and a velocity separation of $246\pm46\; \rm{km\; s^{-1}}$.

By fitting an exponential disk model to the observed velocity field, we showed
 that a simple rotating disk cannot fit the velocity field on its own. Thus, a 
 more complex velocity field is needed, but the SN of the present data does not allow a good 
 constraint to be had. This also implies that obtaining an accurate dynamical mass for the
 8 o'clock system is not possible at present.
 
 Similar to some of clumpy galaxies studied by \citet{genzel11}, the 8 o'clock arc shows a blue-shifted wing but with a less broad profile. We note that as can be seen for example from Fig. \ref{fig:reconstruction-450w}, the galaxy has a very clumpy nature in the source plane, but because of the lack of spatial resolution, we are not able to study these clumps in more detail.

\section*{Acknowledgements}
We are very thankful for useful comments and suggestions of the anonymous referee.
We would like also to thank Ali Rahmati for his useful comments on this paper, 
Raymond Oonk and Benoit Epinat for useful 
discussion about SINFONI data reduction, 
Richard Davies for providing us with the {\scriptsize{DYSMAL}} code, 
Johan Richard for his help on the lens modelling and also 
Max Pettini, Alicia Berciano Alba,Thomas Martinsson and Joanna Holt for useful discussions. 

We would like to express our appreciation to Huan Lin, Michael Strauss, Chris 
Kochanek, Alice Shapley, Dieter Lutz, Chuck Steidel, and Christy Tremonti for their help 
on the $HST$ proposal along with our spacial thanks to Andrew Baker.

M. Sh., S. A. and D. T. acknowledge the support of Mel Ulmer at North Western University for providing 
 them a meeting room and working place in 2011 May-June.

S.V. is grateful to John McKean for useful comments and discussions on the lens modelling.

During part of this work S.V. was supported by a Pappalardo Fellowship at MIT. 

This research has made use of the Interactive Data Language ( {\scriptsize{IDL}}) and 
 {\scriptsize{QFitsView}}\footnote{www.mpe.mpg.de/$\sim$ott/QFitsView}.

\begin{appendix}
\section{Gaussian decomposition}
\label{sec:appendixA}
As we have shown in Section \ref{sec:spatially-resolved}, the resolved \hb\ profiles are not well fitted by a single Gaussian. 
Here we show the best-fit Gaussian intensity maps of the main and blue-shifted components of the galaxy for the image A2 and 
A3 in Fig. \ref{fig:a1} and Fig. \ref{fig:a2}, respectively. As we mentioned earlier, during the fitting we require the lines to have 
the same velocity widths. In both figures, the upper panel shows the Gaussian intensity map of the main component and 
the lower panel shows the Gaussian intensity map of the blue-shifted component. Intensities are in units of $10^{-17} \rm{erg\;cm^{-2}\;s^{-1}}$. 
We see that the main and the blue-shifted components of the galaxy are offset spatially ($\sim$ 1 arcsec) in the A2 image as we see also in Fig. \ref{fig:bgr} showing three calculated SFR maps of the A2 image in blue, green and red channels. However, for the A3 image it is not possible to decompose these components spatially.

\begin{figure}
\centerline{{\includegraphics[width=0.4\textwidth, angle=90]
           {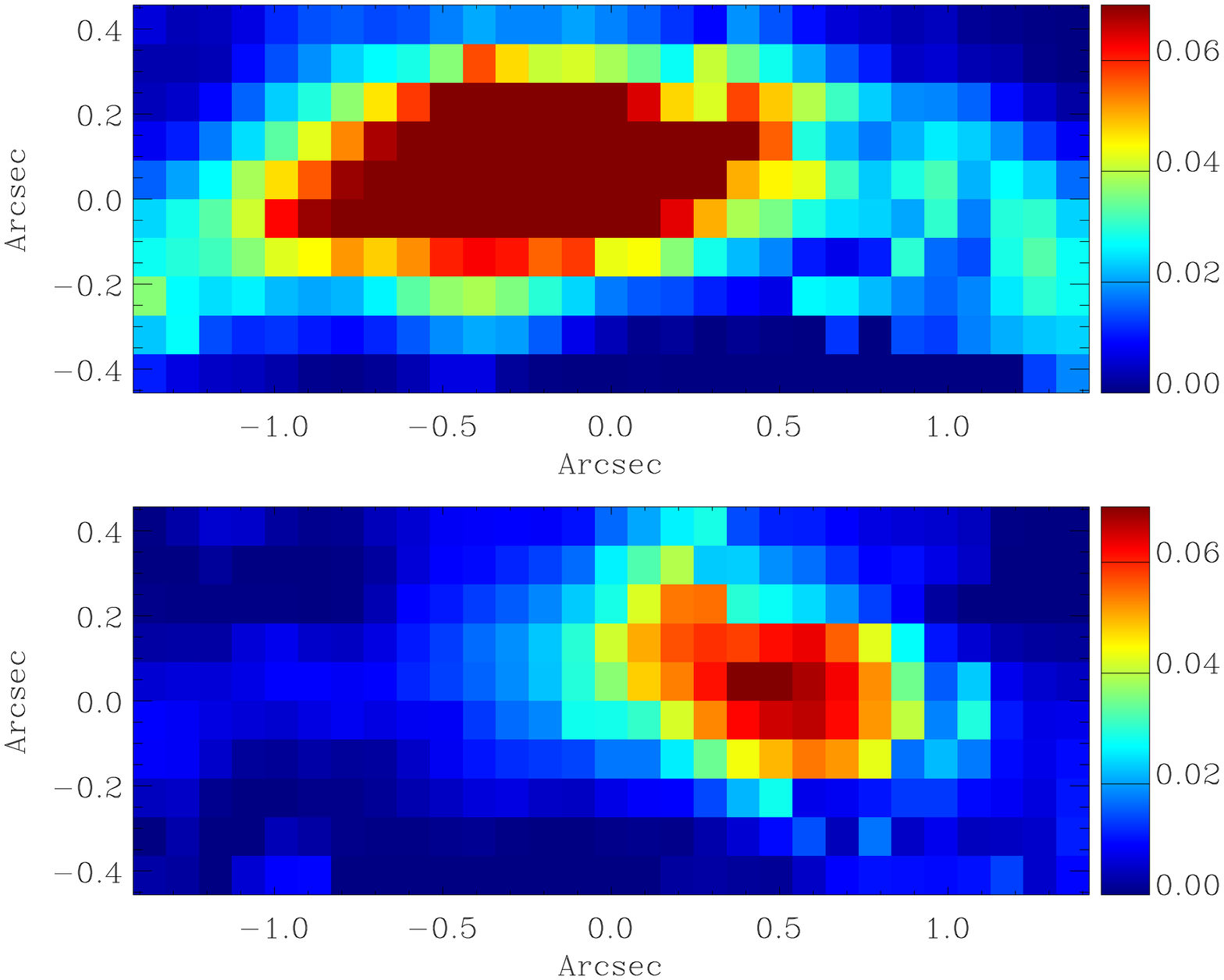}}}
       \caption{Lensed image A2: upper panel shows the Gaussian intensity map of the main component; lower 
       panel shows the Gaussian intensity map of the blue-shifted component. Intensities are in unit of $10^{-17} \rm{erg\;cm^{-2}\;s^{-1}}$.}
\label{fig:a1}
\end{figure}

\begin{figure}
\centerline{{\includegraphics[width=0.5\textwidth, angle=90]
           {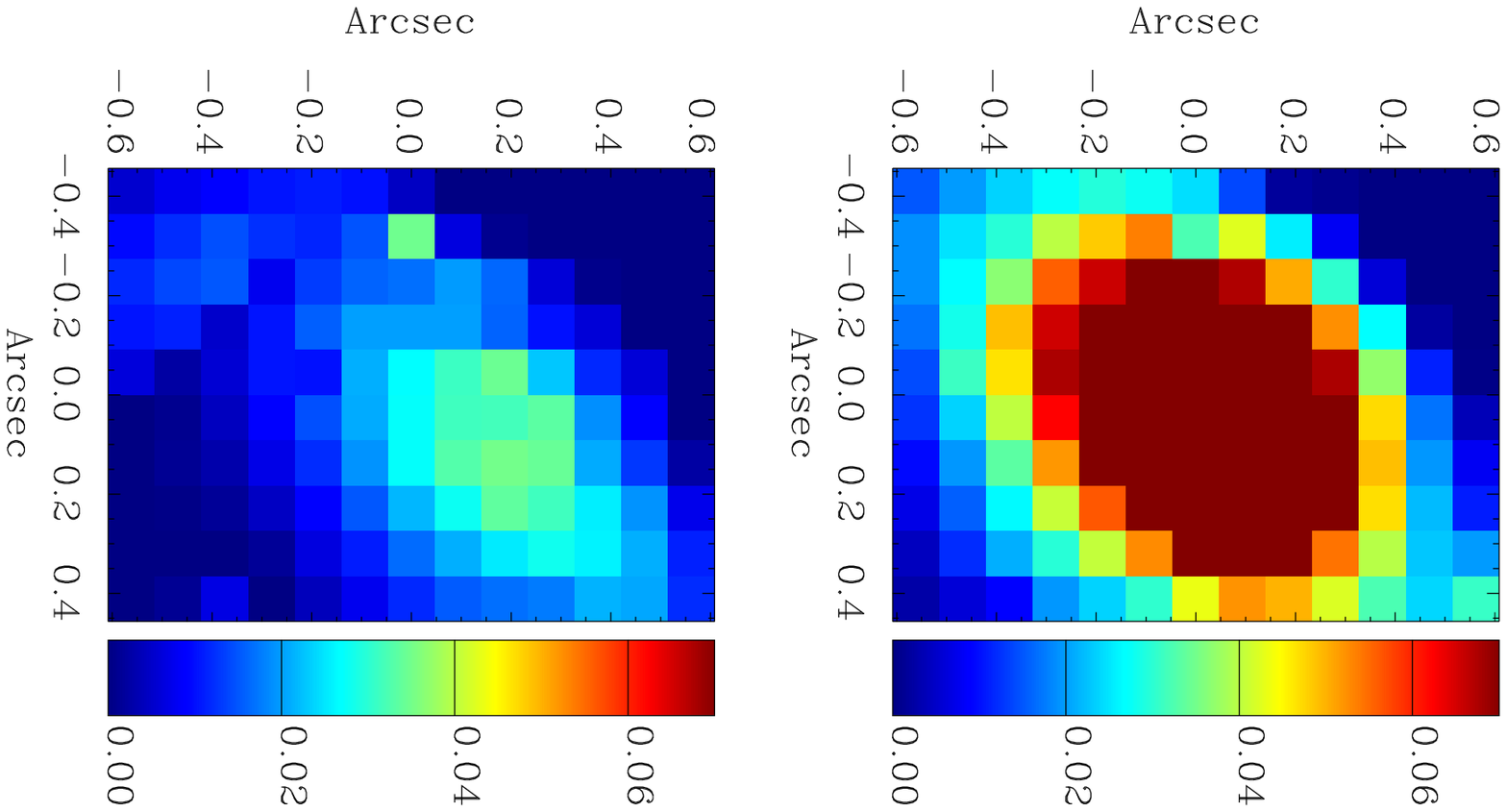}}}
       \caption{Lensed image A3: upper panel shows the Gaussian intensity map of the main component; lower 
       panel shows the Gaussian intensity map of the blue-shifted component. Intensities are in unit of $10^{-17} \rm{erg\;cm^{-2}\;s^{-1}}$. }
\label{fig:a2}
\end{figure}

\section{Outflowing gas}
\label{sec:appendixB}
DZ10 measure significant outflow column densities extending to $-400\; \rm{km \; s^{-1}}$. \hb\ is also detected in those velocities in the outflowing parts (which is the eastern part of both left and right components in image A2, see Fig. \ref{fig:bgr}).
In Fig. \ref{fig:a3} we compare the \hb\ map at $-400 \;\rm{km \; s^{-1}}$ (bottom panel) and the integrated blue part of the \hb\ (4855 \AA-4859 \AA\ or from $-400 \;\rm{km \; s^{-1}}$ to $-140\; \rm{km \; s^{-1}}$). From this figure we see that \hb\ is detected in the outflowing part with 3 $\sigma$.

\begin{figure}
\centerline{{\includegraphics[trim=5cm 1cm 0cm 0cm, clip=true,width=0.5\textwidth, angle=90]
           {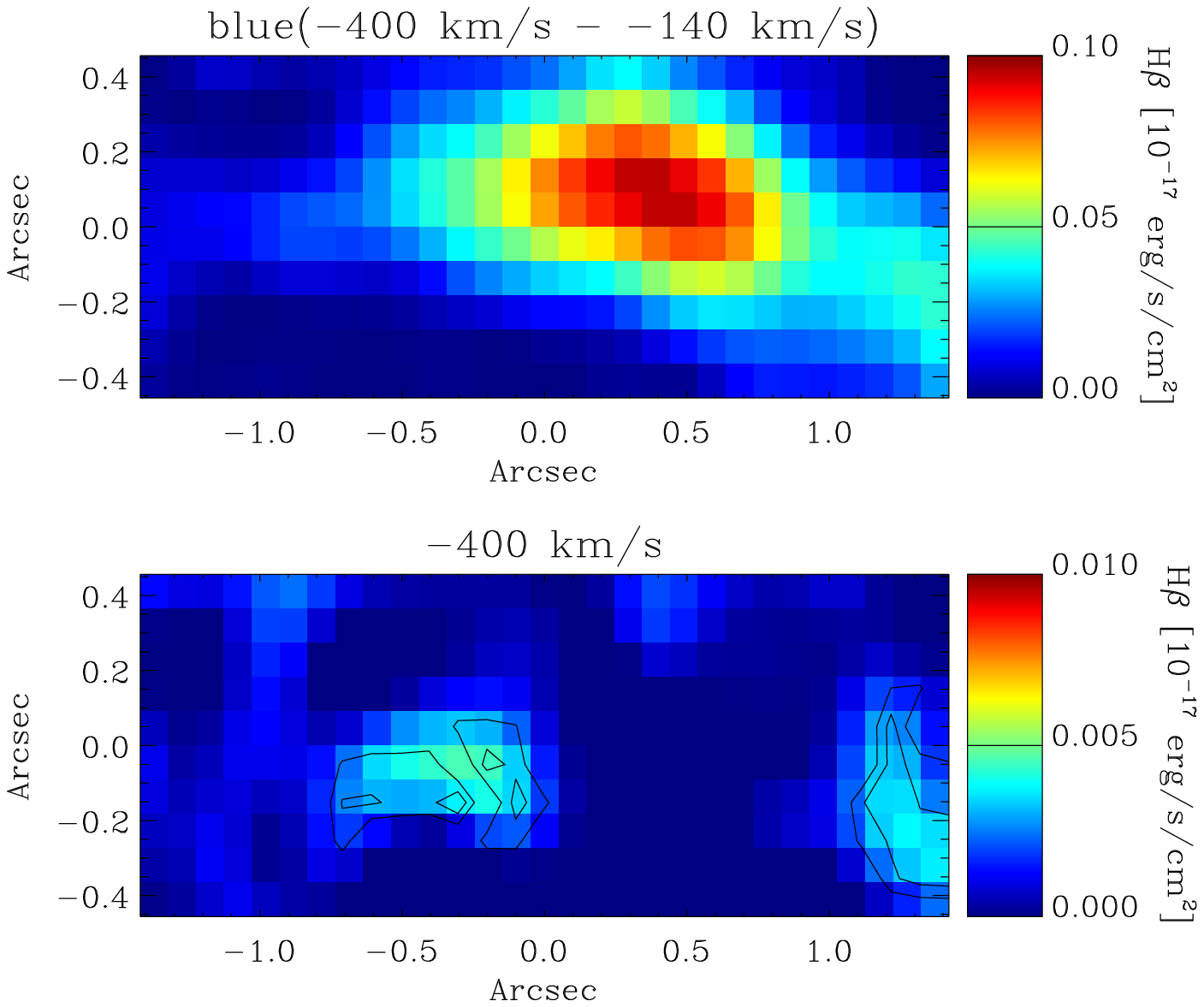}}}
       \caption{Lensed image A2: upper panel shows the \hb\ map integrated over (4855 \AA-4859 \AA\ or from $-400\; \rm{km \; s^{-1}}$ to $-140 \;\rm{km \; s^{-1}}$); lower 
       panel shows the \hb\ detected at $-400 \;\rm{km \; s^{-1}}$. Intensities are in unit of $10^{-17} \rm{erg\;cm^{-2}\;s^{-1}}$. From this figure we see that \hb\ is detected in $-400\; \rm{km \; s^{-1}}$ in the outflowing part. Contours show 3$\sigma$ and 5$\sigma$ significance.}
\label{fig:a3}
\end{figure}

\end{appendix}
\end{document}